\documentclass[preprintnumbers,amsmath,amssymb,superscriptaddress,twocolumn,showpacs]{revtex4-2}
\usepackage{graphicx}
\usepackage{dcolumn}
\usepackage{bm}
\usepackage{physics}
\usepackage[caption=false]{subfig}

\newcommand{\be}{\begin{equation}}
\newcommand{\ee}{\end{equation}}
\newcommand{\bea}{\begin{eqnarray}}
\newcommand{\eea}{\end{eqnarray}}

\begin{document}

\title{Spin-Relaxation of Dipolar-Coupled Nitrogen-Vacancy Centers  : \\ The role of Double-flip Processes}

\author{C. Pellet-Mary$^1$, M. Perdriat$^1$, P. Huillery$^2$,  G. H\'etet} 

\affiliation{Laboratoire De Physique de l'\'Ecole Normale Sup\'erieure, \'Ecole Normale Sup\'erieure, PSL Research University, CNRS, Sorbonne Universit\'e, Universit\'e Paris Cit\'e , 24 rue Lhomond, 75231 Paris Cedex 05, France \\$^2$Univ Rennes, INSA Rennes, CNRS, Institut FOTON - UMR 6082, F-35000 Rennes, France}

\begin{abstract}
We study relaxation processes of the spins of dipolar-coupled negatively charged nitrogen vacancy (NV$^-$) centers under transverse electric fields and magnetic fields. Specifically, we uncover regimes where flip-flop, double-flip processes as well as mixing induced by local electric fields play a significant role in NV-NV cross-relaxation.
Our results are relevant for understanding decoherence in many-body spin systems as well as for high sensitivity magneto- and electro-metry with long-lived interacting solid-state spins. As a proof of principle, we present an orientation and microwave-free magnetometer based on cross-relaxation.
\end{abstract}

\maketitle

The electronic spin properties of the negatively charged nitrogen-vacancy (NV$^-$) center in diamond have given rise to a wealth of applications in nanoscale sensing and quantum information science thanks in part to the possibility to optically polarize and read-out its spin state under ambient conditions \cite{DOHERTY20131}. In particular, ensembles of NV centers are widely studied for their enhanced magnetic field sensing capabilities \cite{Acosta, TALLAIRE2020421,edmonds2021characterisation, chatzidrosos2021fiberized, Barry, Bauch} and as pristine platforms for observing many-body effects \citep{kucsko2018critical, ChoiNat, ZuYao, dwyer2021probing}. When the NV spin concentration reaches ppm values, spin depolarisation, or cross-relaxation (CR) takes place through a very rich many-body dynamics associated with disorder \citep{choi_observation_2017}. These mechanisms limit the efficiency of typical microwave based NV magnetometers, but quantum control techniques can be combined to surpass this interaction limit \citep{zhou2020quantum}. Further, CR mechanisms can be turned as a tool for magnetic field sensing \cite{akhmedzhanov_microwave-free_2017, akhmedzhanov_magnetometry_2019}. 
Spectral features in the photoluminescence (PL) indeed appear when the magnetic field crosses specific crystal planes where dipolar interactions are enhanced, leading to CR. The projected sensitivity of such magnetometers lie in the tens of pT$/\sqrt{\rm Hz}$  \cite{akhmedzhanov_microwave-free_2017}, on a par with the most sensitive microwave based NV magnetometers \cite{Wolf, Sturner}. 

Cross-relaxation features close to zero magnetic field have also been observed and could be deployed for higher sensitivity microwave-free magnetometry \cite{filimonenko2018weak, filimonenko2022manifestation}.
The CR contrast was indeed shown to be much larger in this zero-field limit \citep{jarmola_longitudinal_2015,  mrozek_longitudinal_2015} but all the relaxation mechanisms have not been identified there.
Here, we study dipolar relaxation processes in ensembles of NV centers in the presence of small transverse electric and magnetic fields. Specifically, by employing magnetic field scans along specific crystalline directions, we find regimes where flip-flop, double-flip processes as well as mixing induced by local electric fields play a role. We also present an orientation and microwave-free magnetometer that employs double-flip processes in the dipolar interaction between NV centers.

\begin{figure}
\includegraphics[width=0.45\textwidth]{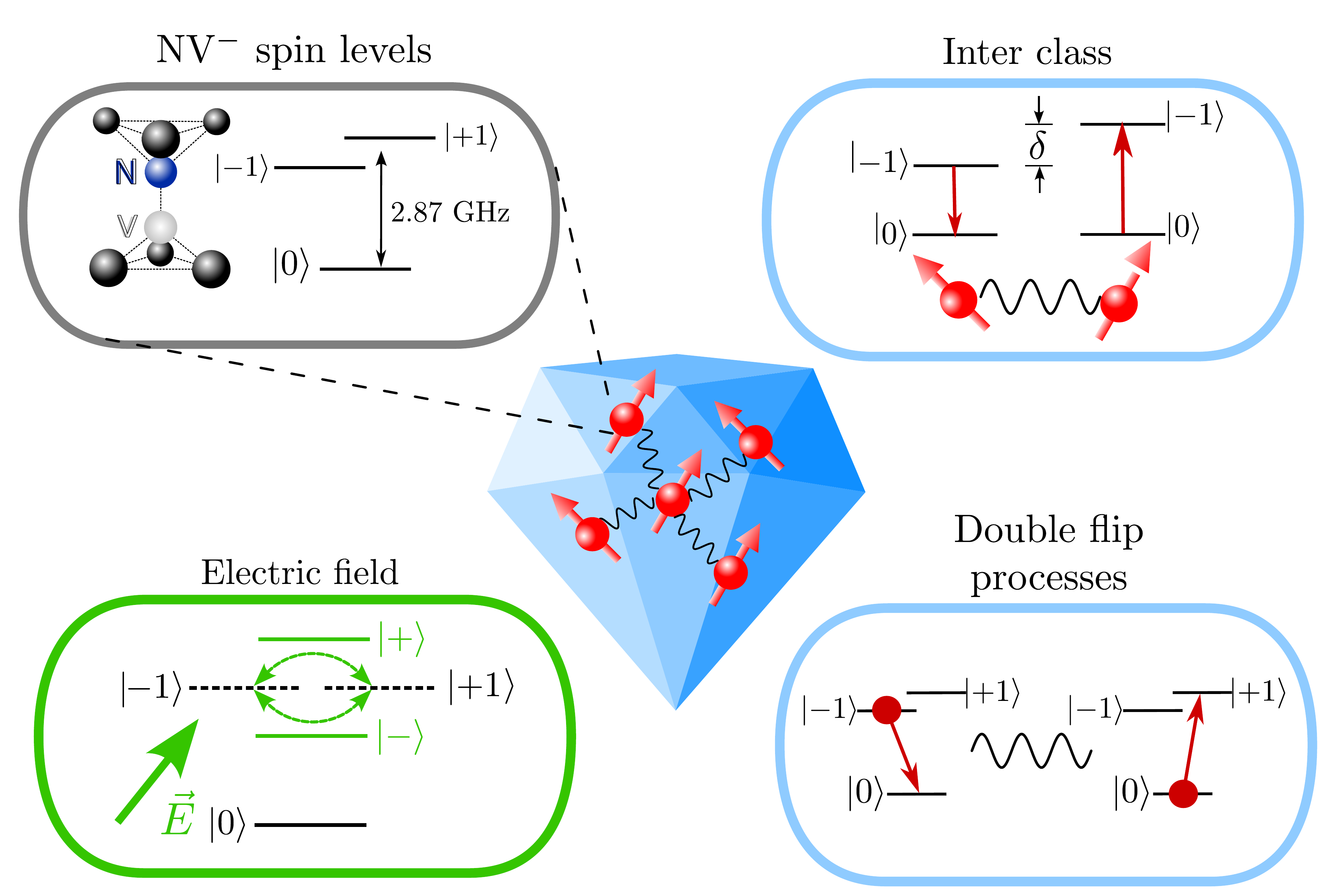}
\caption{Schematics showing a diamond with interacting spins (central picture) as well as three processes that account for dipolar relaxation: flip-flop processes involving two different classes of NV centers, local electric field mixing and double-flip processes where two units of spin angular momentum are exchanged. }
\label{schema_intro}
\end{figure}

The electronic spin of NV center is a spin-1 system in the electronic ground state (see Fig. 1, top left panel), which can be optically polarized in the $\ket{m_s=0}$ state. The PL of this state is also larger than in the $\ket{m_s=\pm 1}$ states enabling spin-read out at room temperature \cite{DOHERTY20131}. The $\ket{m_s=\pm 1}$ spin states are separated from the $\ket{m_s=0}$ state by  $D = (2\pi) 2.87$ GHz so that when a resonant microwave or static transverse magnetic field is applied, the PL is reduced \citep{epstein2005anisotropic,lai2009influence}.  The processes that lead to depolarization in strongly coupled spin-1 systems are depicted in Fig.~1. Flip-flop processes involving coupling between spins with identical or different orientations, or ``classes", are depicted in the top-right panel. They have already been shown to play a dominant role in dense ensembles of NV centers when $B \gtrsim 30$~G : tuning the difference $\delta$ between the different NV centers' spin states indeed result in a strong $T_1$ reduction \cite{choi_observation_2017}.  We will show here that double-flip up and down as well as mixing induced by local electric fields (cf. two bottom panels) can also give a significant contribution to the spin depolarization close to zero-field.

\begin{figure}
\includegraphics[width=0.45\textwidth]{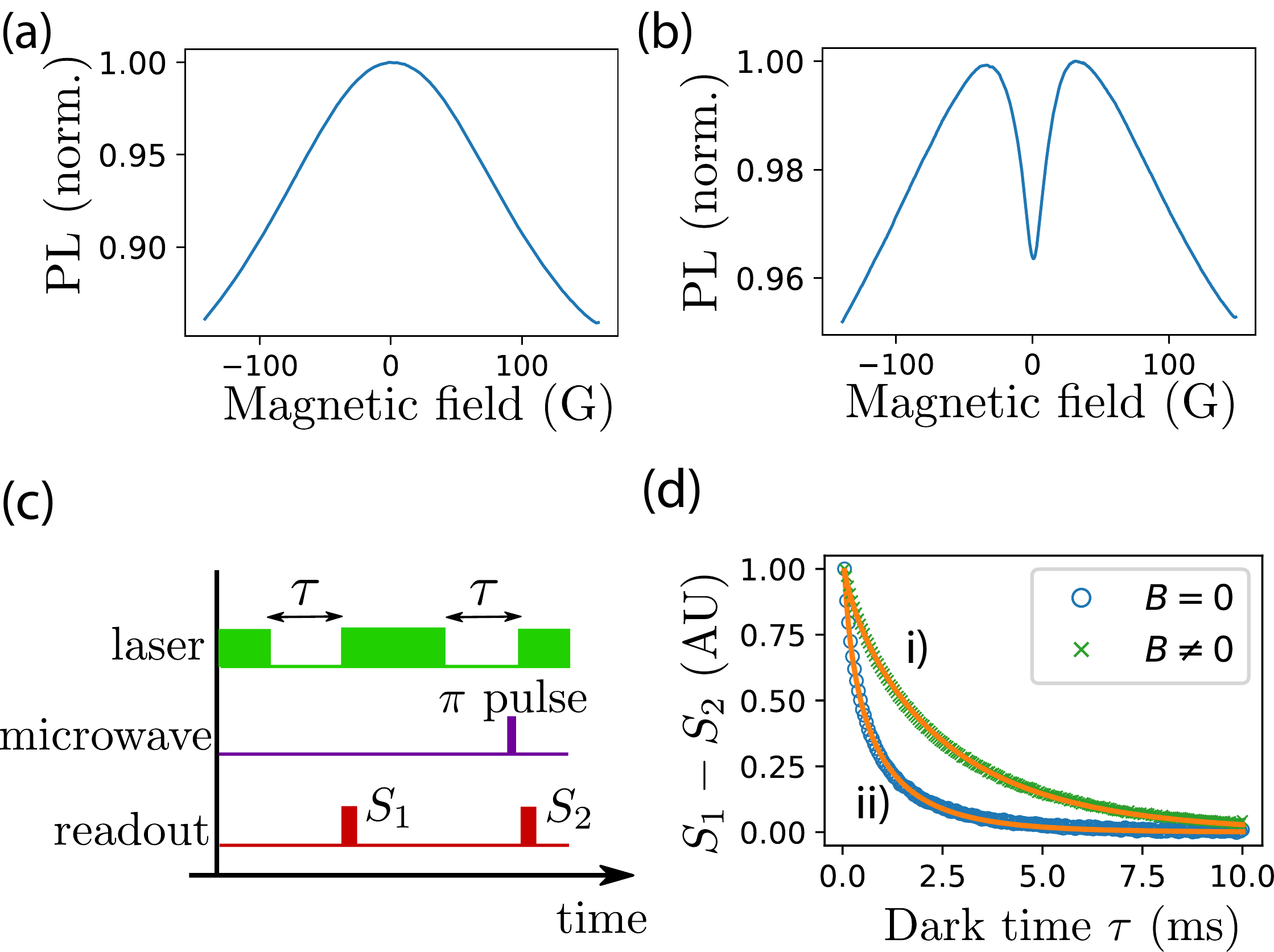}
\caption{a) and b) : Photoluminescence from NV center ensembles as a function of a magnetic field applied in an arbitrary direction for sample CVD-1 with [NV$^-$]$\approx 50\ \rm ppb$ and for sample HPHT-150-1 with [NV$^-$]$\approx 3\ \rm ppm$ respectively. (c) Sequence used to measure the spin lifetime. (d) Trace i) and ii) are the spin relaxation signals $S_1-S_2$ measured for the dense sample at zero and non-zero magnetic fields respectively. The fitting procedure (orange plain line) is described in the main text.}
\label{T1}
\end{figure}

The experimental apparatus and the samples used in this study together with their nomenclature are presented in the supplementary material (SM) sec. II and III \citep{SI_low_filed_CR}\nocite{anishchik2015low, filimonenko2020weak, van1990electric}. 
Fig.~\ref{T1}-a) and b) show the change in PL as a function of a magnetic field applied {\it via} a homebuilt electromagnet for two diamond samples with a low and high concentration of NV$^-$ centers respectively (CVD-1 and HPHT-150-1). In both samples, we see a decrease in PL as the magnetic field amplitude increases. There is however a stark difference in the low magnetic field region where only high-density samples shows a drop in PL \citep{jarmola_longitudinal_2015,  mrozek_longitudinal_2015}. 

\begin{figure*}
\includegraphics[width=.95\textwidth]{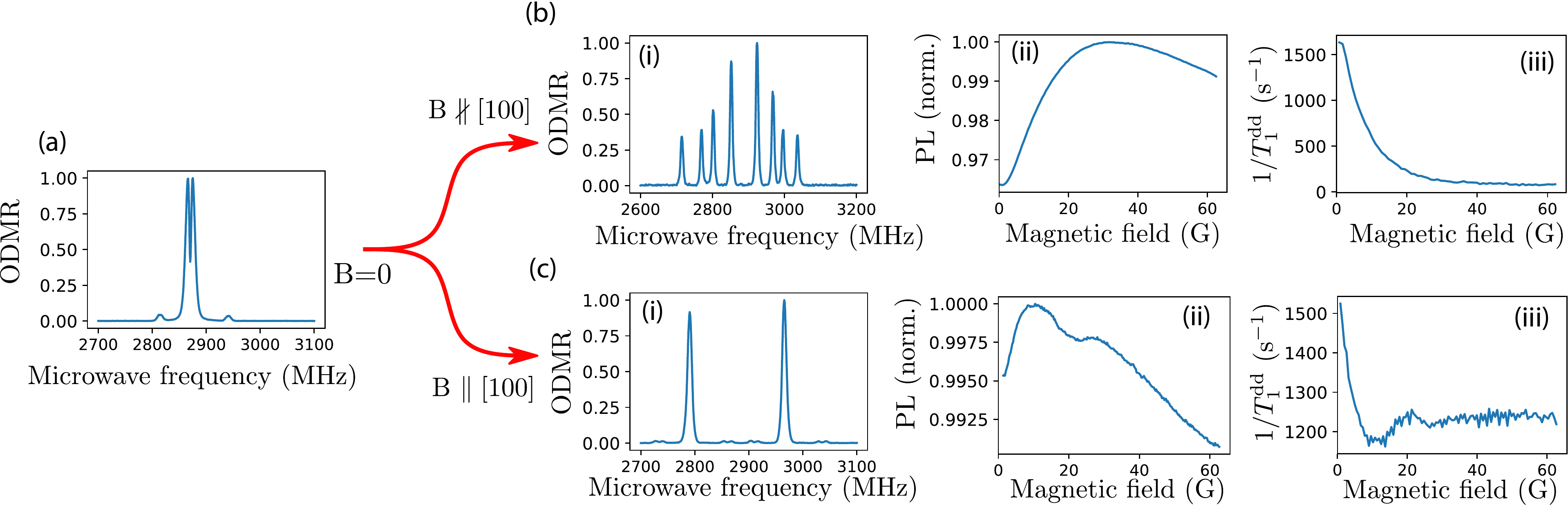}
\caption{(a) ODMR spectrum in zero field with microwave amplitude modulation (see experimental details in SM sec. III). (b)-i) ODMR spectrum for a magnetic field $\approx$ 60 G, misaligned by $\sim  24^\circ$ from the [100] axis. (b)-ii) Normalized PL of the NV$^-$ ensemble as a function of the magnetic field amplitude. (b)-iii) Stretched part of the spin decay $1/T_1^{\rm dd}$ as a function of the magnetic field amplitude. (c)-i), (c)-ii) and (c)-iii): same measurements as (b) but with a magnetic field close to the [100] axis. All the measurements were realized using the sample HPHT-150-1.}
\label{100_VS_1x4}
\end{figure*}
This effect has been attributed to CR between the NV centers through dipole-dipole coupling \citep{mrozek_longitudinal_2015, choi2017depolarization}. 
We will denote $T_1^{\rm ph}$ the characteristic timescale of phonon-induced relaxation and $T_1^{\rm dd}$ the density dependent timescale associated with dipole-dipole  interactions.
Fig.~\ref{T1}-c) shows the sequence employed to measure $T_1$ and Fig.~\ref{T1}-d) shows the result of the measurement when $B\approx 0$ and $B \approx 50\ \rm G$.
All dense samples used in this study show $T_1^{\rm dd} \sim T_1^{\rm ph}$ (see SM sec. IV A), so both decay processes are included in the analysis. Fitting the two traces of Fig.~2 d) by the product of a simple and a stretched exponential, we find $T_1^{\rm ph}=3.6$ ms for both curves and $T_1^{\rm dd}= 0.6\ \rm ms$ and $13.0\ \rm ms$ for trace i) and ii) respectively. These results thus demonstrates a twenty-fold increase in the dipolar depolarization rate when the B field is turned to zero.
The fluctuator model developed in \citep{choi2017depolarization} contains some of the explanation for this increase in dipolar depolarization: all four NV classes in the diamond are resonant in zero field, which increases the flip-flop rates. However, the model did not include extra specificities in the zero magnetic field region, namely the role of local electric field and double flip processes. 


Fig.~\ref{100_VS_1x4}-a) and b)-i) show optically detected magnetic resonance (ODMR) spectra with and without magnetic field, from the dense ensemble HPHT-150-1. From the resonance line-widths in a magnetic field, we extract decoherence rates $T_2^*$ in the hundreds of nanosecond range, limited by the coupling between NV centers and the fluctuating spins of substitutional nitrogen atoms (also called $P_1$ centers).
Using the same magnetic field alignment, Fig.~\ref{100_VS_1x4} (b)-ii) and iii) show the PL and the $1/ T_1^{\rm dd}$ decay rates from the NV center ensemble as a function of the amplitude of the magnetic field. 
We observe that the 4~\% increase in PL as the magnetic field increases is indeed correlated with a drop in the spin decay rate from 1600~s$^{-1}$ to 80~s$^{-1}$. This result can again be explained by the decreasing number of resonant spins as the zero magnetic field is increased. We also confirmed that the half width of the dip ($\approx 10$~G) is consistent with a fluctuator model taking into account the reduction of the two spin resonances overlap as the $B$ field is increased (see SM sec. IV C).

Fig.~\ref{100_VS_1x4} (c)-i) instead shows an ODMR where all four classes are brought to resonance by placing the magnetic field along the [100] crystalline axis. Fig.~\ref{100_VS_1x4} (c)-ii) and (c)-iii) show the change in the PL and in $1/T_1^{\rm dd}$ as a function of a magnetic field that is aligned in this direction. Surprisingly, we can still observe an increase in the spin decay rate and a corresponding drop in the PL when the magnetic field tends to zero, although all classes are always resonant (see SM, sec. IV-E).
We also note that the PL and $1/T_1^{\rm dd}$ change is also much sharper in this scenario.
Note that the slight drop in PL and the corresponding bump for $1/T_1^{\rm dd}$ at $B \sim 20$G is related to dipolar interaction with NV centers that have a $^{13}$C as a first neighbor \cite{pellet2021optical}. 

When $\gamma | \mathbf{B} |\gg 1/T_2^*$, the only resonant terms in the dipole-dipole coupling between NV centers are the flip-flop terms $~\ket{0,\pm 1}\bra{\pm 1,0}$ (see SM sec. V).
In zero magnetic field however, other resonant mechanisms of the dipolar Hamiltonian have to be taken into consideration, which may elucidate the above result.
It was shown in  \cite{mittiga2018imaging} that local electric fields coming from $P_1^+$ or NV$^-$ centers are responsible for ODMR profiles at zero magnetic field. 
In order to estimate their contributions in the spin relaxation, we perform numerical simulations where we add the following Hamiltonian for the electric field dependent NV$^-$ ground state: 
\begin{equation}
\label{NV Hamiltonian}
\mathcal{H}_{\rm elec}/h= d_\perp \left[ E_x(\hat S_y^2-\hat S_x^2) + E_y(\hat S_x \hat S_y+\hat S_y \hat S_x) \right].
\end{equation}
$E_{x,y}$ are the projections of local electric fields on the NV axes and $d_\perp=17~$Hz $\cdot$ cm/V is the transverse electric field susceptibility. 
Under local electric fields with orientations given by the angle $\phi_E={\rm tan}(E_x/E_y)$, the eigenstates of the total hamiltonian of the NV spins in the ground state are $\ket{0}$ and $\ket{\pm}=\frac{1}{\sqrt{2}}(\ket{+1}\pm e^{-i\phi_E}\ket{-1})$.
Computing the influence of the flip-flop terms $\ket{\pm,0} \bra{0,\pm} $ in the fluctuator model shows that $1/T_1^{\rm dd}$ indeed increases at zero field, even after averaging over all possible angles $\phi_E$ (see SM, sec. V-E).  Another mechanism at stake when considering dipolar interactions between spin-1 systems at low fields are double-flip processes. 
These processes correspond to the terms $\ket{+1,0} \bra{0,-1}$ in the dipolar interaction, giving rise to an exchange of two units of spin-angular momentum when the $\ket{\pm 1}$ states are resonant. Simulations (see SM, sec V-F) suggest that double-flips contribute significantly to the spin relaxation at low fields.

\begin{figure}
\includegraphics[width=0.45\textwidth]{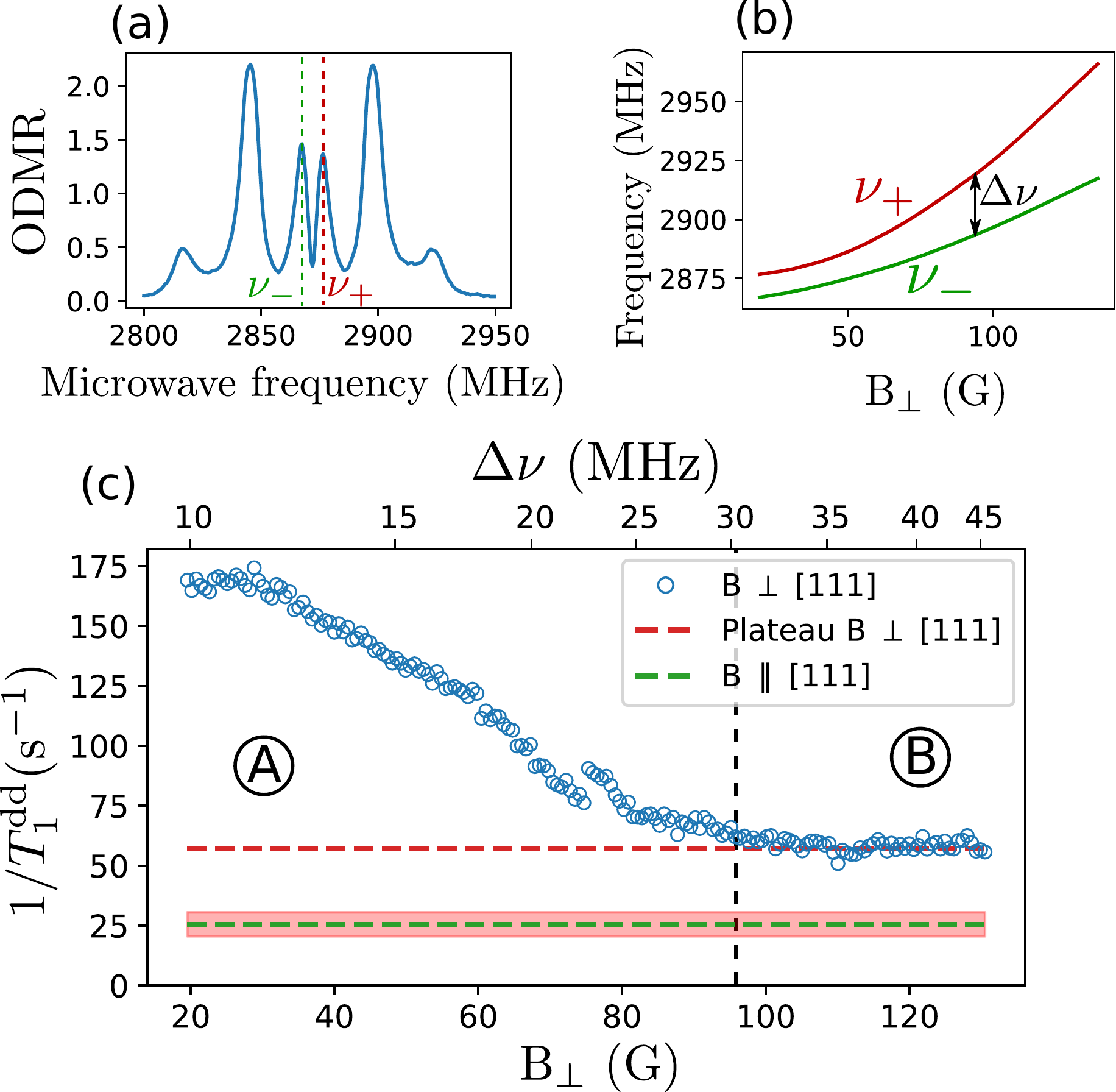}
\caption{(a) ODMR spectrum for $B_\perp=20\ \rm G$. The two transition frequencies of the class that is orthogonal to $B$ are denoted $\nu_+$ and $\nu_-$. (b) Measurement of $\nu_+$ and $\nu_-$ through ODMR as a function of $B_\perp$. (c) Measurement of $1/T_1^{\rm dd}$ for a single NV class as a function of $B_\perp$. The red dashed line is the value reached for high transverse field. The green dashed line correspond to the value found for the same sample and NV class, under a longitudinal magnetic field. Error bars are depicted by the shaded pink region. The detuning $\Delta \nu = \nu_+-\nu-$ is indicated on the top $x$-axis. Vertical black dashed line indicates the separation between regions A and B (see main text).}
\label{B_transverse}
\end{figure}

To quantify these two processes experimentally, we apply a purely transverse magnetic field on one of the NV class. Indeed, when $B_{\perp} \leq 150\ \rm G$, the eigenstates of the spin Hamiltonian are close to $\ket{0}$, $\ket{+}$ and $\ket{-}$ to $\approx 2\%$ (see SM sec. IV-F). We can thus use the transverse field to emulate the role of the electric field, a property that has previously been used to increase the electric field sensing ability of NV centers \cite{dolde2011electric,qiu2022nanoscale}.
Fig.~\ref{B_transverse} (a) shows an ODMR spectrum where $B_{\perp}=20\ \rm G$ on sample HPHT-150-2. The central two lines correspond to the $\ket{0}\to \ket{-}$ and $\ket{0}\to \ket{+}$ transitions for the class that is orthogonal to $\bm B$.
Fig. \ref{B_transverse}-(c) shows the measurement of the decay rate $1/T_1^{\rm dd}$ for that class, as a function of $B_{\perp}$. Two regions can be observed on this graph: in region A the decay rate decreases with $B_{\perp}$, while in region B it stabilizes to a value $1/T_1^{\rm dd}=60\pm 5\ \rm{s}^{-1}$. We also indicated the value $1/T_1^{\rm dd}=25\pm 5\ \rm{s}^{-1}$ found for the same class but employing a longitudinal magnetic field. 

\begin{figure}[ht]
\includegraphics[width=0.45\textwidth]{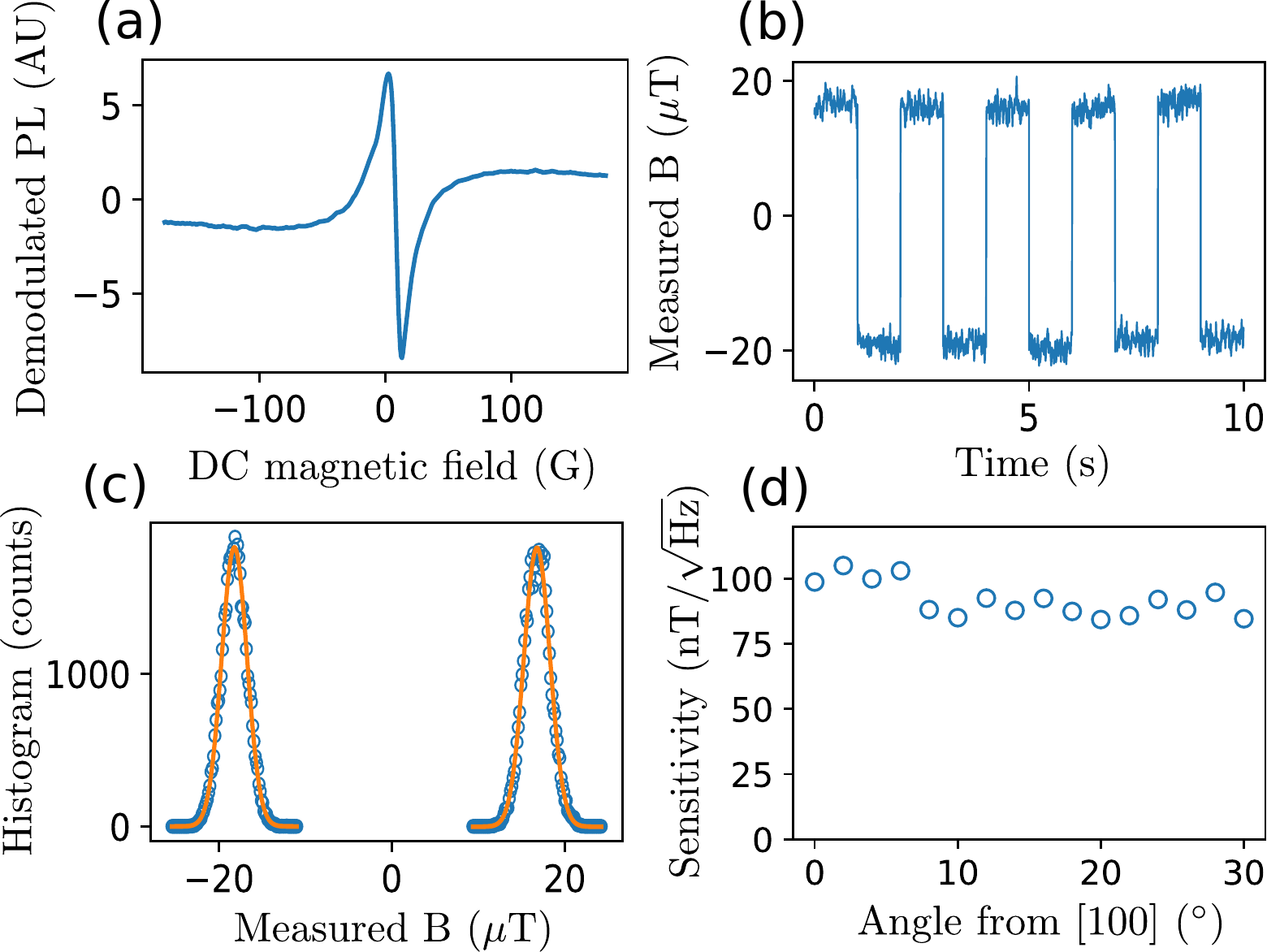}
\caption{Low-field magnetometry protocol. (a) Demodulated PL as a function of an externally applied magnetic field. (b) External magnetic field read-out. (c) Histogram of the measurement in Fig. (b) fitted by gaussians of standard deviation $\sigma=1.5\ \mu \rm T$. (d) Measured sensitivity as function of the angle between the external magnetic field and the [100] crystalline axis.}
\label{magneto}
\end{figure}

To understand these results, we plot the detuning $\Delta \nu$ between the states $\ket{+}$ and $\ket{-}$ as a function of $B_{\perp}$ in Fig. \ref{B_transverse} (b). When $\delta \nu > 1/T_2^*$, these states do not overlap any longer so double-flips are efficient in region A, while in region B only flip-flops in the basis $\ket{\pm}$ play a role. We note that the decay rate in the $\{ \ket{0}, \ket{\pm} \}$ basis is more than a factor of two larger than the decay rate coming from intra-class flip-flops in the $\{\ket{0}, \ket{\pm 1} \}$ basis. This effect is corroborated by our model detailed in SM sec. V. Importantly, in region A, the double-flips processes give a maximum decay rate that is $\sim 5$ times greater than the decay caused by the change of basis alone. From these results and according to our model, the double-flips are thus the dominant cause of zero-field depolarization observed in Fig. \ref{100_VS_1x4} (c)-iii).

Our observations have important implications for magnetometry with NV ensembles.
DC microwave-free magnetometry has already been performed using either NV-NV cross-relaxations \citep{akhmedzhanov_microwave-free_2017,akhmedzhanov_magnetometry_2019} or level anti-crossing \citep{Wickenbrock, zheng2017level, zheng_microwave-free_2020}. Here we propose a similar protocol but using the spin depolarization at zero-field. 
To do so, we use a lock-in detection and add a magnetic field modulation at $\sim 1\ \rm kHz$ with an amplitude $\sim 10\ \rm G$ through the same electromagnet. Fig. \ref{magneto}-(a) shows the demodulated PL while a DC magnetic field is scanned in an arbitrary direction. Here, we use the sample HPHT-15-1 with a laser power $\sim 1\ \rm mW$. The optical power in the collected PL is $\sim 1\ \mu\rm W$.  We can see a sharp linear slope in low field $\abs{B} < 5\ \rm G$. Once calibrated, the slope provides a 1D magnetometer, which could be extended to 3D with a set of 3 coils or 3 electromagnets, as in \cite{zheng_microwave-free_2020}. In order to assess the sensitivity of the measurement, we alternate a small DC field of $\approx 40\ \mu\rm{T}$ every few seconds and take a histogram of the measured fields (with a total duration of $\approx $50 s), as shown in Fig. \ref{magneto}-(b) and (c). The histogram is well fitted by gaussians of standard deviation $\sigma=1.5\ \mu \rm T$. The measurement was performed here with an output low-pass filter of time constant $\tau=3\ \rm ms$, which gives us a DC sensitivity $\eta=\sigma \sqrt{\tau}=82\ n\rm{T}/\sqrt{\rm Hz}$. This value corresponds to $\eta/\sqrt{V}\approx 4.7 \ \mu \rm{T}/\mu m^{3/2}/\sqrt{\rm Hz}$ when normalized to the volume. Importantly, this measurement is consistent with the experimentally found $\sim 5\ \mu \rm{T}/\sqrt{\rm Hz}$ sensitivity obtained with sample HPHT-1-1 (see samples details in SM sec. II).  Our results thus offer bright prospects for high-sensitivity magnetometry using NV centers in a millimeter size volumes. 

Compared to previously employed protocols, the sensitivity does not crucially depend on crystalline orientation, making this magnetometer principle operational with diamond powders or polycrystalline samples. Another significant advantage is that this measurement is insensitive to thermal fluctuations and strain inhomogeneities because dipolar-relaxation at zero-field does not depend on the zero-field splitting. In other typical NV magnetometers however, the zero-field splitting dependence with temperature and strain can yield systematic errors \cite{Bauch, Barry}. 

We now evaluate the relative role of the three causes of spin depolarization that assist the operation of the magnetometer. To do so, we measure its sensitivity while changing the angle of the magnetic field.  The results are shown in Fig.~\ref{magneto} (d). We observe only a $\sim$ 10\% improvement in the sensitivity as we leave the [100] region.
When $\bm B \parallel [100]$ only the double flips and the electric field cause a depolarization, whereas the three effects are at play in all the other orientations. It should be noted that this observation is sample dependent. Other samples, including those from the same batch, have shown a higher orientation dependence, corresponding to a lower contribution from local electric fields and double-flip processes.
Interestingly, the double-flips and electric field effects are here the dominant factors in the sensitivity of this protocol. 

As a conclusion, we identified three mechanisms causing spin depolarization in zero field for dense ensemble of NV$^-$ centers, all related to an increase in the dipole-dipole induced CR between the spins of NV centers. The lift in degeneracy of the spin state of different NV classes was found to be the main cause of zero-field depolarization, followed by double-flip processes and then the electric field induced mixing. We have employed CR for microwave-free as well as for orientation-free DC-magnetometry and demonstrated sensitivities in the $\eta/\sqrt{V}\approx 5 \ \mu \rm{T}/\mu m^{3/2}/\sqrt{\rm Hz}$ range for commercially available diamonds and show that that double-flips and local electric fields play a critical role. 
We believe that our results will also be important for microwave-based low field magnetometry \cite{Vetter_LFM, WangRB},  as well as for understanding many-body phenomena \citep{kucsko2018critical, ChoiNat, ZuYao, dwyer2021probing} and spin-mechanical effects \cite{pellet2021magnetic} with strongly coupled spins. \\

{\bf Acknowledgments}
\\
We would like to acknowledge support from Alexandre Tallaire and Jocelyn Achard as well as 
SIRTEQ for funding.

\bibliographystyle{plain}
\bibliography{CRmain}{}

\begin{thebibliography}{10}

\bibitem{SI_low_filed_CR}
See supplemental material at [...], which includes refs. [20-22].

\bibitem{Acosta}
V.~M. Acosta, E.~Bauch, M.~P. Ledbetter, C.~Santori, K.-M.~C. Fu, P.~E.
  Barclay, R.~G. Beausoleil, H.~Linget, J.~F. Roch, F.~Treussart,
  S.~Chemerisov, W.~Gawlik, and D.~Budker.
\newblock Diamonds with a high density of nitrogen-vacancy centers for
  magnetometry applications.
\newblock {\em Phys. Rev. B}, 80:115202, Sep 2009.

\bibitem{akhmedzhanov_microwave-free_2017}
Rinat Akhmedzhanov, Lev Gushchin, Nikolay Nizov, Vladimir Nizov, Dmitry
  Sobgayda, Ilya Zelensky, and Philip Hemmer.
\newblock Microwave-free magnetometry based on cross-relaxation resonances in
  diamond nitrogen-vacancy centers.
\newblock {\em Phys. Rev. A}, 96(1):013806, July 2017.
\newblock Number: 1.

\bibitem{akhmedzhanov_magnetometry_2019}
Rinat Akhmedzhanov, Lev Gushchin, Nikolay Nizov, Vladimir Nizov, Dmitry
  Sobgayda, Ilya Zelensky, and Philip Hemmer.
\newblock Magnetometry by cross-relaxation-resonance detection in ensembles of
  nitrogen-vacancy centers.
\newblock {\em Phys. Rev. A}, 100(4):043844, October 2019.
\newblock Number: 4.

\bibitem{anishchik2015low}
SV~Anishchik, VG~Vins, AP~Yelisseyev, NN~Lukzen, NL~Lavrik, and VA~Bagryansky.
\newblock Low-field feature in the magnetic spectra of nv- centers in diamond.
\newblock {\em New Journal of Physics}, 17(2):023040, 2015.

\bibitem{Barry}
John~F. Barry, Jennifer~M. Schloss, Erik Bauch, Matthew~J. Turner, Connor~A.
  Hart, Linh~M. Pham, and Ronald~L. Walsworth.
\newblock Sensitivity optimization for nv-diamond magnetometry.
\newblock {\em Rev. Mod. Phys.}, 92:015004, Mar 2020.

\bibitem{Bauch}
Erik Bauch, Connor~A. Hart, Jennifer~M. Schloss, Matthew~J. Turner, John~F.
  Barry, Pauli Kehayias, Swati Singh, and Ronald~L. Walsworth.
\newblock Ultralong dephasing times in solid-state spin ensembles via quantum
  control.
\newblock {\em Phys. Rev. X}, 8:031025, Jul 2018.

\bibitem{chatzidrosos2021fiberized}
Georgios Chatzidrosos, Joseph~Shaji Rebeirro, Huijie Zheng, Muhib Omar, Andreas
  Brenneis, Felix~M St{\"u}rner, Tino Fuchs, Thomas Buck, Robert R{\"o}lver,
  Tim Schneemann, et~al.
\newblock Fiberized diamond-based vector magnetometers.
\newblock {\em Frontiers in Photonics}, page~4, 2021.

\bibitem{choi2017depolarization}
Joonhee Choi, Soonwon Choi, Georg Kucsko, Peter~C Maurer, Brendan~J Shields,
  Hitoshi Sumiya, Shinobu Onoda, Junichi Isoya, Eugene Demler, Fedor Jelezko,
  et~al.
\newblock Depolarization dynamics in a strongly interacting solid-state spin
  ensemble.
\newblock {\em Physical review letters}, 118(9):093601, 2017.

\bibitem{ChoiNat}
Soonwon Choi, Joonhee Choi, Renate Landig, Georg Kucsko, Hengyun Zhou, Junichi
  Isoya, Fedor Jelezko, Shinobu Onoda, Hitoshi Sumiya, Vedika Khemani, Curt von
  Keyserlingk, Norman~Y. Yao, Eugene Demler, and Mikhail~D. Lukin.
\newblock Observation of discrete time-crystalline order in a disordered
  dipolar many-body system.
\newblock {\em Nature}, 543(7644):221--225, 2017.

\bibitem{choi_observation_2017}
Soonwon Choi, Joonhee Choi, Renate Landig, Georg Kucsko, Hengyun Zhou, Junichi
  Isoya, Fedor Jelezko, Shinobu Onoda, Hitoshi Sumiya, Vedika Khemani, Curt von
  Keyserlingk, Norman~Y. Yao, Eugene Demler, and Mikhail~D. Lukin.
\newblock Observation of discrete time-crystalline order in a disordered
  dipolar many-body system.
\newblock {\em Nature}, 543(7644):221--225, March 2017.
\newblock Number: 7644 Publisher: Nature Publishing Group.

\bibitem{DOHERTY20131}
Marcus~W. Doherty, Neil~B. Manson, Paul Delaney, Fedor Jelezko, Jörg
  Wrachtrup, and Lloyd~C.L. Hollenberg.
\newblock The nitrogen-vacancy colour centre in diamond.
\newblock {\em Physics Reports}, 528(1):1 -- 45, 2013.
\newblock The nitrogen-vacancy colour centre in diamond.

\bibitem{dolde2011electric}
Florian Dolde, Helmut Fedder, Marcus~W Doherty, Tobias N{\"o}bauer, Florian
  Rempp, Gopalakrishnan Balasubramanian, Thomas Wolf, Friedemann Reinhard,
  Lloyd~CL Hollenberg, Fedor Jelezko, et~al.
\newblock Electric-field sensing using single diamond spins.
\newblock {\em Nature Physics}, 7(6):459--463, 2011.

\bibitem{dwyer2021probing}
Bo~L Dwyer, Lila~VH Rodgers, Elana~K Urbach, Dolev Bluvstein, Sorawis
  Sangtawesin, Hengyun Zhou, Yahia Nassab, Mattias Fitzpatrick, Zhiyang Yuan,
  Kristiaan De~Greve, et~al.
\newblock Probing spin dynamics on diamond surfaces using a single quantum
  sensor.
\newblock {\em arXiv preprint arXiv:2103.12757}, 2021.

\bibitem{edmonds2021characterisation}
Andrew~M Edmonds, Connor~A Hart, Matthew~J Turner, Pierre-Olivier Colard,
  Jennifer~M Schloss, Kevin~S Olsson, Raisa Trubko, Matthew~L Markham, Adam
  Rathmill, Ben Horne-Smith, et~al.
\newblock Characterisation of cvd diamond with high concentrations of nitrogen
  for magnetic-field sensing applications.
\newblock {\em Materials for Quantum Technology}, 1(2):025001, 2021.

\bibitem{epstein2005anisotropic}
RJ~Epstein, FM~Mendoza, YK~Kato, and DD~Awschalom.
\newblock Anisotropic interactions of a single spin and dark-spin spectroscopy
  in diamond.
\newblock {\em Nature physics}, 1(2):94--98, 2005.

\bibitem{filimonenko2022manifestation}
DS~Filimonenko, VM~Yasinskii, Alexander~P Nizovtsev, S~Ya Kilin, and Fedor
  Jelezko.
\newblock Manifestation in ir-luminescence of cross relaxation processes
  between nv-centers in weak magnetic fields.
\newblock {\em Journal of Applied Spectroscopy}, 88(6):1131--1143, 2022.

\bibitem{filimonenko2018weak}
DS~Filimonenko, VM~Yasinskii, AP~Nizovtsev, and S~Ya Kilin.
\newblock Weak magnetic field resonance effects in diamond with
  nitrogen-vacancy centers.
\newblock {\em Semiconductors}, 52(14):1865--1867, 2018.

\bibitem{filimonenko2020weak}
DS~Filimonenko, VM~Yasinskii, AP~Nizovtsev, S~Ya Kilin, and Fedor Jelezko.
\newblock Weak magnetic field effects on the photoluminescence of an ensemble
  of nv centers in diamond: experiment and modelling.
\newblock {\em Semiconductors}, 54(12):1730--1733, 2020.

\bibitem{jarmola_longitudinal_2015}
A.~Jarmola, A.~Berzins, J.~Smits, K.~Smits, J.~Prikulis, F.~Gahbauer,
  R.~Ferber, D.~Erts, M.~Auzinsh, and D.~Budker.
\newblock Longitudinal spin-relaxation in nitrogen-vacancy centers in electron
  irradiated diamond.
\newblock {\em Appl. Phys. Lett.}, 107(24):242403, December 2015.
\newblock Number: 24.

\bibitem{kucsko2018critical}
Georg Kucsko, Soonwon Choi, Joonhee Choi, Peter~C Maurer, Hengyun Zhou, Renate
  Landig, Hitoshi Sumiya, Shinobu Onoda, Junich Isoya, Fedor Jelezko, et~al.
\newblock Critical thermalization of a disordered dipolar spin system in
  diamond.
\newblock {\em Physical review letters}, 121(2):023601, 2018.

\bibitem{lai2009influence}
Ngoc~Diep Lai, Dingwei Zheng, Fedor Jelezko, Fran{\c{c}}ois Treussart, and
  Jean-Fran{\c{c}}ois Roch.
\newblock Influence of a static magnetic field on the photoluminescence of an
  ensemble of nitrogen-vacancy color centers in a diamond single-crystal.
\newblock {\em Applied Physics Letters}, 95(13):133101, 2009.

\bibitem{mittiga2018imaging}
Thomas Mittiga, Satcher Hsieh, Chong Zu, Bryce Kobrin, Francisco Machado,
  Prabudhya Bhattacharyya, NZ~Rui, Andrey Jarmola, Soonwon Choi, Dmitry Budker,
  et~al.
\newblock Imaging the local charge environment of nitrogen-vacancy centers in
  diamond.
\newblock {\em Physical review letters}, 121(24):246402, 2018.

\bibitem{mrozek_longitudinal_2015}
Mariusz Mrózek, Daniel Rudnicki, Pauli Kehayias, Andrey Jarmola, Dmitry
  Budker, and Wojciech Gawlik.
\newblock Longitudinal spin relaxation in nitrogen-vacancy ensembles in
  diamond.
\newblock {\em EPJ Quantum Technol.}, 2(1):22, December 2015.
\newblock Number: 1.

\bibitem{pellet2021magnetic}
C~Pellet-Mary, P~Huillery, M~Perdriat, and G~H{\'e}tet.
\newblock Magnetic torque enhanced by tunable dipolar interactions.
\newblock {\em Physical Review B}, 104(10):L100411, 2021.

\bibitem{pellet2021optical}
Cl{\'e}ment Pellet-Mary, Paul Huillery, Maxime Perdriat, A~Tallaire, and
  Gabriel H{\'e}tet.
\newblock Optical detection of paramagnetic defects in diamond grown by
  chemical vapor deposition.
\newblock {\em Physical Review B}, 103(10):L100411, 2021.

\bibitem{qiu2022nanoscale}
Ziwei Qiu, Assaf Hamo, Uri Vool, Tony~X Zhou, and Amir Yacoby.
\newblock Nanoscale electric field imaging with an ambient scanning quantum
  sensor microscope.
\newblock {\em arXiv preprint arXiv:2205.03952}, 2022.

\bibitem{Sturner}
Felix~M. Stürner, Andreas Brenneis, Thomas Buck, Julian Kassel, Robert
  Rölver, Tino Fuchs, Anton Savitsky, Dieter Suter, Jens Grimmel, Stefan
  Hengesbach, Michael Förtsch, Kazuo Nakamura, Hitoshi Sumiya, Shinobu Onoda,
  Junichi Isoya, and Fedor Jelezko.
\newblock Integrated and portable magnetometer based on nitrogen-vacancy
  ensembles in diamond.
\newblock {\em Advanced Quantum Technologies}, 4(4):2000111, 2021.

\bibitem{TALLAIRE2020421}
Alexandre Tallaire, Ovidiu Brinza, Paul Huillery, Tom Delord, Clément
  Pellet-Mary, Robert Staacke, Bernd Abel, Sébastien Pezzagna, Jan Meijer,
  Nadia Touati, Laurent Binet, Alban Ferrier, Philippe Goldner, Gabriel Hetet,
  and Jocelyn Achard.
\newblock High nv density in a pink cvd diamond grown with n2o addition.
\newblock {\em Carbon}, 170:421 -- 429, 2020.

\bibitem{van1990electric}
Eric Van~Oort and Max Glasbeek.
\newblock Electric-field-induced modulation of spin echoes of nv centers in
  diamond.
\newblock {\em Chemical Physics Letters}, 168(6):529--532, 1990.

\bibitem{Vetter_LFM}
Philipp~J. Vetter, Alastair Marshall, Genko~T. Genov, Tim~F. Weiss, Nico
  Striegler, Eva~F. Gro\ss{}mann, Santiago Oviedo-Casado, Javier Cerrillo,
  Javier Prior, Philipp Neumann, and Fedor Jelezko.
\newblock Zero- and low-field sensing with nitrogen-vacancy centers.
\newblock {\em Phys. Rev. Applied}, 17:044028, Apr 2022.

\bibitem{WangRB}
Ning Wang, Chu-Feng Liu, Jing-Wei Fan, Xi~Feng, Weng-Hang Leong, Amit Finkler,
  Andrej Denisenko, J\"org Wrachtrup, Quan Li, and Ren-Bao Liu.
\newblock Zero-field magnetometry using hyperfine-biased nitrogen-vacancy
  centers near diamond surfaces.
\newblock {\em Phys. Rev. Research}, 4:013098, Feb 2022.

\bibitem{Wickenbrock}
Arne Wickenbrock, Huijie Zheng, Lykourgos Bougas, Nathan Leefer, Samer Afach,
  Andrey Jarmola, Victor~M. Acosta, and Dmitry Budker.
\newblock Microwave-free magnetometry with nitrogen-vacancy centers in diamond.
\newblock {\em Applied Physics Letters}, 109(5):053505, 2016.

\bibitem{Wolf}
Thomas Wolf, Philipp Neumann, Kazuo Nakamura, Hitoshi Sumiya, Takeshi Ohshima,
  Junichi Isoya, and J\"org Wrachtrup.
\newblock Subpicotesla diamond magnetometry.
\newblock {\em Phys. Rev. X}, 5:041001, Oct 2015.

\bibitem{zheng2017level}
Huijie Zheng, Georgios Chatzidrosos, Arne Wickenbrock, Lykourgos Bougas, Reinis
  Lazda, Andris Berzins, Florian~Helmuth Gahbauer, Marcis Auzinsh, Ruvin
  Ferber, and Dmitry Budker.
\newblock Level anti-crossing magnetometry with color centers in diamond.
\newblock In {\em Slow Light, Fast Light, and Opto-Atomic Precision Metrology
  X}, volume 10119, pages 115--122. SPIE, 2017.

\bibitem{zheng_microwave-free_2020}
Huijie Zheng, Zhiyin Sun, Georgios Chatzidrosos, Chen Zhang, Kazuo Nakamura,
  Hitoshi Sumiya, Takeshi Ohshima, Junichi Isoya, Jörg Wrachtrup, Arne
  Wickenbrock, and Dmitry Budker.
\newblock Microwave-{Free} {Vector} {Magnetometry} with {Nitrogen}-{Vacancy}
  {Centers} along a {Single} {Axis} in {Diamond}.
\newblock {\em Phys. Rev. Applied}, 13(4):044023, April 2020.
\newblock Number: 4.

\bibitem{zhou2020quantum}
Hengyun Zhou, Joonhee Choi, Soonwon Choi, Renate Landig, Alexander~M Douglas,
  Junichi Isoya, Fedor Jelezko, Shinobu Onoda, Hitoshi Sumiya, Paola
  Cappellaro, et~al.
\newblock Quantum metrology with strongly interacting spin systems.
\newblock {\em Physical review X}, 10(3):031003, 2020.

\bibitem{ZuYao}
C.~Zu, F.~Machado, B.~Ye, S.~Choi, B.~Kobrin, T.~Mittiga, S.~Hsieh,
  P.~Bhattacharyya, M.~Markham, D.~Twitchen, A.~Jarmola, D.~Budker, C.~R.
  Laumann, J.~E. Moore, and N.~Y. Yao.
\newblock Emergent hydrodynamics in a strongly interacting dipolar spin
  ensemble.
\newblock {\em Nature}, 597(7874):45--50, 2021.

\end{thebibliography}


\begin{thebibliography}{10}

\bibitem{anishchik2015low}
SV~Anishchik, VG~Vins, AP~Yelisseyev, NN~Lukzen, NL~Lavrik, and VA~Bagryansky.
\newblock Low-field feature in the magnetic spectra of nv- centers in diamond.
\newblock {\em New Journal of Physics}, 17(2):023040, 2015.

\bibitem{choi_depolarization_2017}
Joonhee Choi, Soonwon Choi, Georg Kucsko, Peter~C. Maurer, Brendan~J. Shields,
  Hitoshi Sumiya, Shinobu Onoda, Junichi Isoya, Eugene Demler, Fedor Jelezko,
  Norman~Y. Yao, and Mikhail~D. Lukin.
\newblock Depolarization {Dynamics} in a {Strongly} {Interacting}
  {Solid}-{State} {Spin} {Ensemble}.
\newblock {\em Phys. Rev. Lett.}, 118(9):093601, March 2017.
\newblock Number: 9.

\bibitem{filimonenko2020weak}
DS~Filimonenko, VM~Yasinskii, AP~Nizovtsev, S~Ya Kilin, and Fedor Jelezko.
\newblock Weak magnetic field effects on the photoluminescence of an ensemble
  of nv centers in diamond: experiment and modelling.
\newblock {\em Semiconductors}, 54(12):1730--1733, 2020.

\bibitem{giri_selective_2019}
R.~Giri, C.~Dorigoni, S.~Tambalo, F.~Gorrini, and A.~Bifone.
\newblock Selective measurement of charge dynamics in an ensemble of
  nitrogen-vacancy centers in nanodiamond and bulk diamond.
\newblock {\em Phys. Rev. B}, 99(15):155426, April 2019.
\newblock Number: 15.

\bibitem{giri_coupled_2018}
R.~Giri, F.~Gorrini, C.~Dorigoni, C.~E. Avalos, M.~Cazzanelli, S.~Tambalo, and
  A.~Bifone.
\newblock Coupled charge and spin dynamics in high-density ensembles of
  nitrogen-vacancy centers in diamond.
\newblock {\em Phys. Rev. B}, 98(4):045401, July 2018.
\newblock Number: 4.

\bibitem{Hall}
L.~T. Hall, P.~Kehayias, D.~A. Simpson, A.~Jarmola, A.~Stacey, D.~Budker, and
  L.~C.~L. Hollenberg.
\newblock Detection of nanoscale electron spin resonance spectra demonstrated
  using nitrogen-vacancy centre probes in diamond.
\newblock {\em Nature Communications}, 7(1):10211, 2016.

\bibitem{jarmola_temperature-_2012}
A.~Jarmola, V.~M. Acosta, K.~Jensen, S.~Chemerisov, and D.~Budker.
\newblock Temperature- and {Magnetic}-{Field}-{Dependent} {Longitudinal} {Spin}
  {Relaxation} in {Nitrogen}-{Vacancy} {Ensembles} in {Diamond}.
\newblock {\em Phys. Rev. Lett.}, 108(19):197601, May 2012.
\newblock Number: 19.

\bibitem{mittiga2018imaging}
Thomas Mittiga, Satcher Hsieh, Chong Zu, Bryce Kobrin, Francisco Machado,
  Prabudhya Bhattacharyya, NZ~Rui, Andrey Jarmola, Soonwon Choi, Dmitry Budker,
  et~al.
\newblock Imaging the local charge environment of nitrogen-vacancy centers in
  diamond.
\newblock {\em Physical review letters}, 121(24):246402, 2018.

\bibitem{mrozek_longitudinal_2015}
Mariusz Mrózek, Daniel Rudnicki, Pauli Kehayias, Andrey Jarmola, Dmitry
  Budker, and Wojciech Gawlik.
\newblock Longitudinal spin relaxation in nitrogen-vacancy ensembles in
  diamond.
\newblock {\em EPJ Quantum Technol.}, 2(1):22, December 2015.
\newblock Number: 1.

\bibitem{TALLAIRE2020421}
Alexandre Tallaire, Ovidiu Brinza, Paul Huillery, Tom Delord, Clément
  Pellet-Mary, Robert Staacke, Bernd Abel, Sébastien Pezzagna, Jan Meijer,
  Nadia Touati, Laurent Binet, Alban Ferrier, Philippe Goldner, Gabriel Hetet,
  and Jocelyn Achard.
\newblock High nv density in a pink cvd diamond grown with n2o addition.
\newblock {\em Carbon}, 170:421 -- 429, 2020.

\bibitem{van1990electric}
Eric Van~Oort and Max Glasbeek.
\newblock Electric-field-induced modulation of spin echoes of nv centers in
  diamond.
\newblock {\em Chemical Physics Letters}, 168(6):529--532, 1990.

\end{thebibliography}

\end{document}


\vspace{0.2in}
{\Large \hspace{1.6in}\textsc{Supplementary Material} }\\
\
\title{Spin-Relaxation of Dipolar-Coupled Nitrogen-Vacancy Centers  : \\ The role of Double-flip Processes}

\author{C. Pellet-Mary$^1$, M. Perdriat$^1$, P. Huillery$^2$,  G. H\'etet$^1$} 

\affiliation{$^1$ Laboratoire De Physique de l'\'Ecole Normale Sup\'erieure, \'Ecole Normale Sup\'erieure, PSL Research University, CNRS, Sorbonne Universit\'e, Universit\'e Paris Cit\'e , 24 rue Lhomond, 75231 Paris Cedex 05, France \\ $^2$ Univ Rennes, INSA Rennes, CNRS, Institut FOTON - UMR 6082, F-35000 Rennes, France}

\maketitle

\tableofcontents

\section{NV$^-$ ground state Hamiltonian under low magnetic field}
\label{sec Hamiltonian}
\begin{figure}[h]
\includegraphics[width=0.9\textwidth]{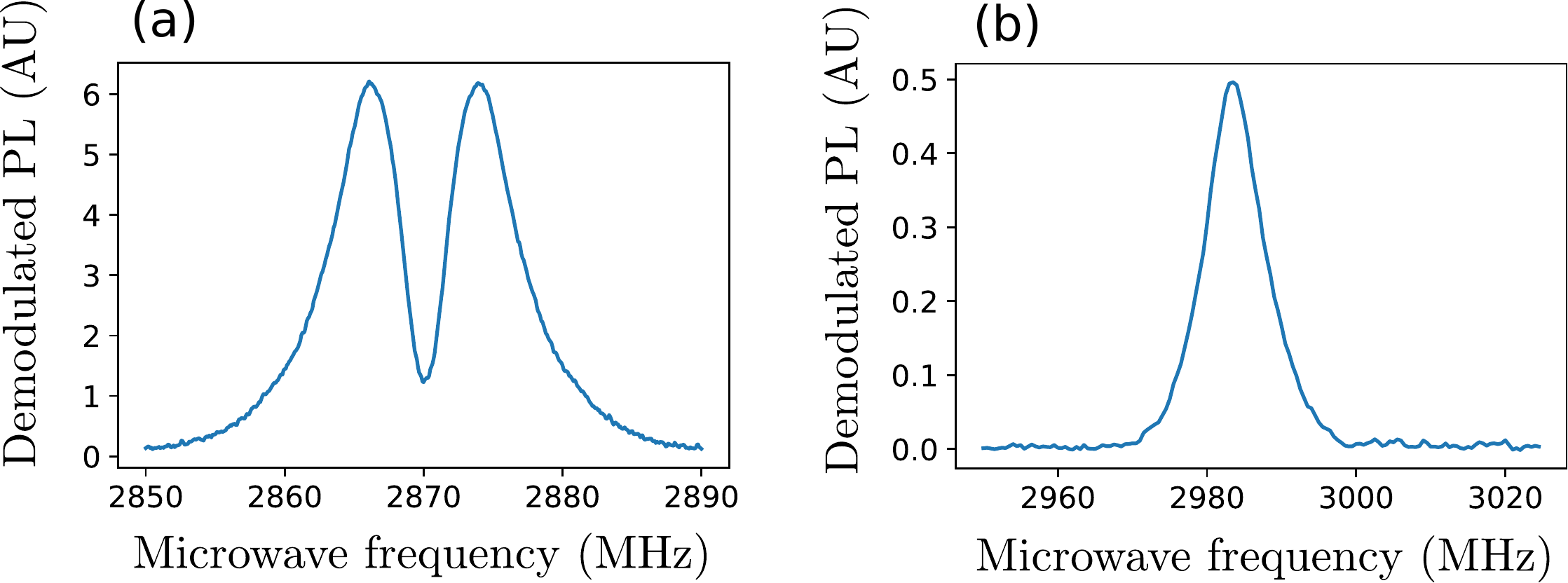}
\caption{ODMR measurement (a) under zero magnetic field, (b) under a magnetic field of $~ 50 G$ when zooming on a single NV class.}
\label{ESR_single_spin}
\end{figure}
Under zero external magnetic field, there are three possible processes that can cause a splitting of the $\{\ket{+1},\ket{-1}\}$ states : local electric fields, crystal strain and local magnetic fields. Of these three causes, only the electric field can explain the shape of the ODMR line that is observed under zero external magnetic field (Fig. \ref{ESR_single_spin}a) \cite{mittiga2018imaging}.

Indeed, random local magnetic field would produce a single broadened line while crystal strain would shift the zero field splitting (ZFS) by the same order of magnitude as the splitting between the levels, which would merge the two transitions onto a single line. Due to the large difference between the longitudinal and transverse electric field susceptibilities ($d_\parallel = 0.35\ \rm Hz\, cm/V$ and $d_\perp = 17\ \rm Hz\, cm/V$ \cite{van1990electric}), random local electric field can, on average, cause a splitting that is much stronger than the ZFS shift and result in a two peak spectrum centered on 2.87 GHz.

\begin{figure}[h]
\includegraphics[width=\textwidth]{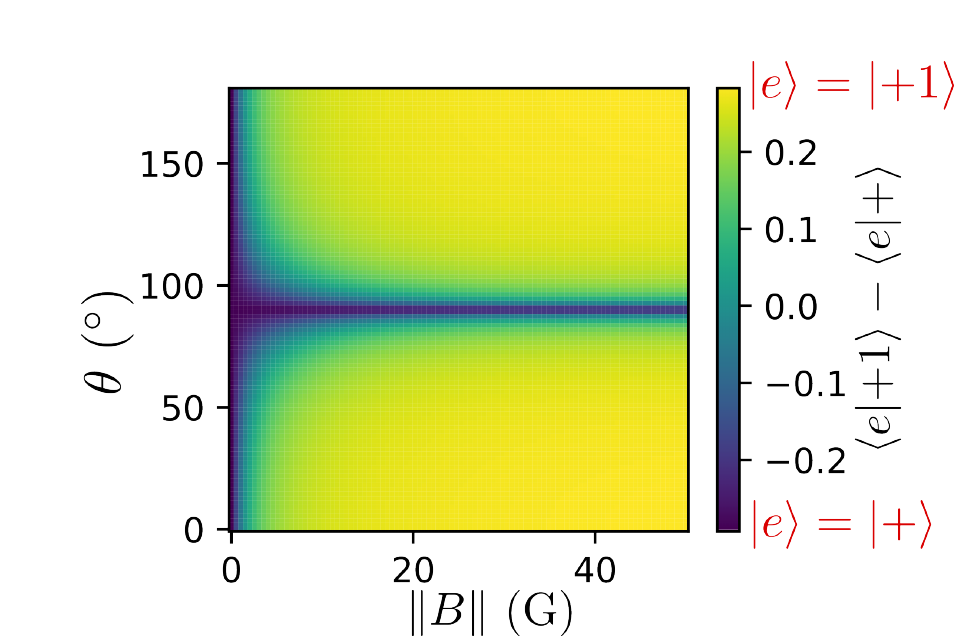}
\caption{Numerical simulations showing the closeness of the Hamiltonian second excited state $\ket{e}$ with the states $\ket{+1}$ and $\ket{+}$, as a function of the magnetic field amplitude and angle $\theta$ with respect to the NV axis. A value of $d_\perp E_\perp = 4\ \rm MHz$ was chosen.}
\label{map etats propres}
\end{figure}

In our model for  dipole-dipole coupling under small magnetic fields, we will therefore neglect the contribution of the strain, local magnetic fields and longitudinal electric fields.  We also do not take into account the hyper-fine structure of the NV center because of the large inhomogeneous broadening of the transitions in all our samples (Fig. \ref{ESR_single_spin}b).
We will then consider the following spin Hamiltonian for the NV$^-$ ground state : 
\begin{equation}
\label{NV Hamiltonian}
\mathcal{H}_s/h=D S_z^2 + \gamma_e \bm{B}_{\rm ext} \cdot \bm{S}+ d_\perp \left[ E_x(S_y^2-S_x^2) + E_y(S_xS_y+S_yS_x) \right],
\end{equation}
where $D=2.87\ \rm GHz$ is the zero field splitting and $\gamma_e=2.8\ \rm MHz/G$ the gyromagnetic ratio of the electron.

In the absence of an external magnetic field, the symmetry of the Hamiltonian in the ($xy$) plane allows us to pick the $x$ direction along the electric field. The eigenstates of $\mathcal{H}_s$ then become $\{ \ket{0},\ket{+}=\frac{\ket{+1}+\ket{-1}}{\sqrt{2}},\ket{-}=\frac{\ket{+1}-\ket{-1}}{\sqrt{2}} \} $.

In the presence of a magnetic field at an angle $\theta$ with respect to the NV axis, we denote the Hamiltonian eigenstates $\{ \ket{g},\ket{d}, \ket{e} \} $ in ascending order of energy. Fig. \ref{map etats propres} shows how close the $\ket{e}$ state is to the $\ket{+1}$ and $\ket{+}$ states as a function of the external magnetic field and angle $\theta$. A similar analysis for the  $\ket{d}$, $\ket{-1}$ and $\ket{-}$ states show similar results, while $\ket{g}$ is pretty much equal to $\ket{0}$ for $B<100\ \rm G$.

This result tells us that, in most cases, the $\{ \ket{0},\ket{+},\ket{-} \}$ basis is the good eigen-basis for magnetic fields smaller than a few Gauss, except in the case of pure transverse magnetic field where the $\{ \ket{0},\ket{+},\ket{-} \}$ basis remains a good basis even for large magnetic fields.

\section{Samples}
Here are the various samples used in this study :
\begin{itemize}
\item \textbf{HPHT-150-1} : A 150 $\mu$m HPHT 1b diamond irradiated and annealed to reach a concentration [NV$^-$] $\approx 3\ \rm ppm$. This sample was bought from Adamas Nanotechnology (MDNV150umHi). It is used in Fig. 2 (b), 2(d) and 3 in the main text, as well as Fig. \ref{ESR_single_spin}, \ref{T1_fits} and \ref{alphas} of the SI.
\item \textbf{HPHT-150-2} : Another sample from the same batch as HPHT-150-1. This sample is used in Fig. 4 in the main text and Fig. \ref{largeur_fluct} (c) of the SI.
\item \textbf{HPHT-150-3} : Another sample from the same batch as HPHT-150-1. This sample is used in Fig. \ref{Pola} of the SI.
\item \textbf{HPHT-15-1}  : A 15 $\mu$m diamond with similar properties as the last ones, also bought from Adamas Nanotechnology (MDNV15umHi). This sample is used in Fig. 5 in the main text.
\item \textbf{HPHT-15-2}  : Another sample from the same batch as HPHT-15-1. This sample is used in Fig. \ref{121 VS 22 fig} and \ref{Various ODMR} of the SI.
\item \textbf{HPHT-1-1}  : A 1 $\mu$m diamond with similar properties as the last ones, also bought from Adamas Nanotechnology (MDNV1umHi). This sample is used to perform a sensitivity measurement similar to the one done in Fig. 5 in the main text.
\item \textbf{CVD-1} : A CVD type IIa bulk diamond. This sample has not been irradiated and contains [NV$^-$] $\approx 50\ \rm ppb$. This sample is used in Fig. 2(a) in the main text.
\item \textbf{CVD-2} : A CVD bulk diamond described in \citep{TALLAIRE2020421}. This one has been irradiated and annealed and contains [NV$^-$] $\approx 4\ \rm ppm$. This sample is used in Fig. \ref{Alignment} of the SI.

\end{itemize}

\section{Experimental Setup}
\begin{figure}[h]
\includegraphics[width=0.7\textwidth]{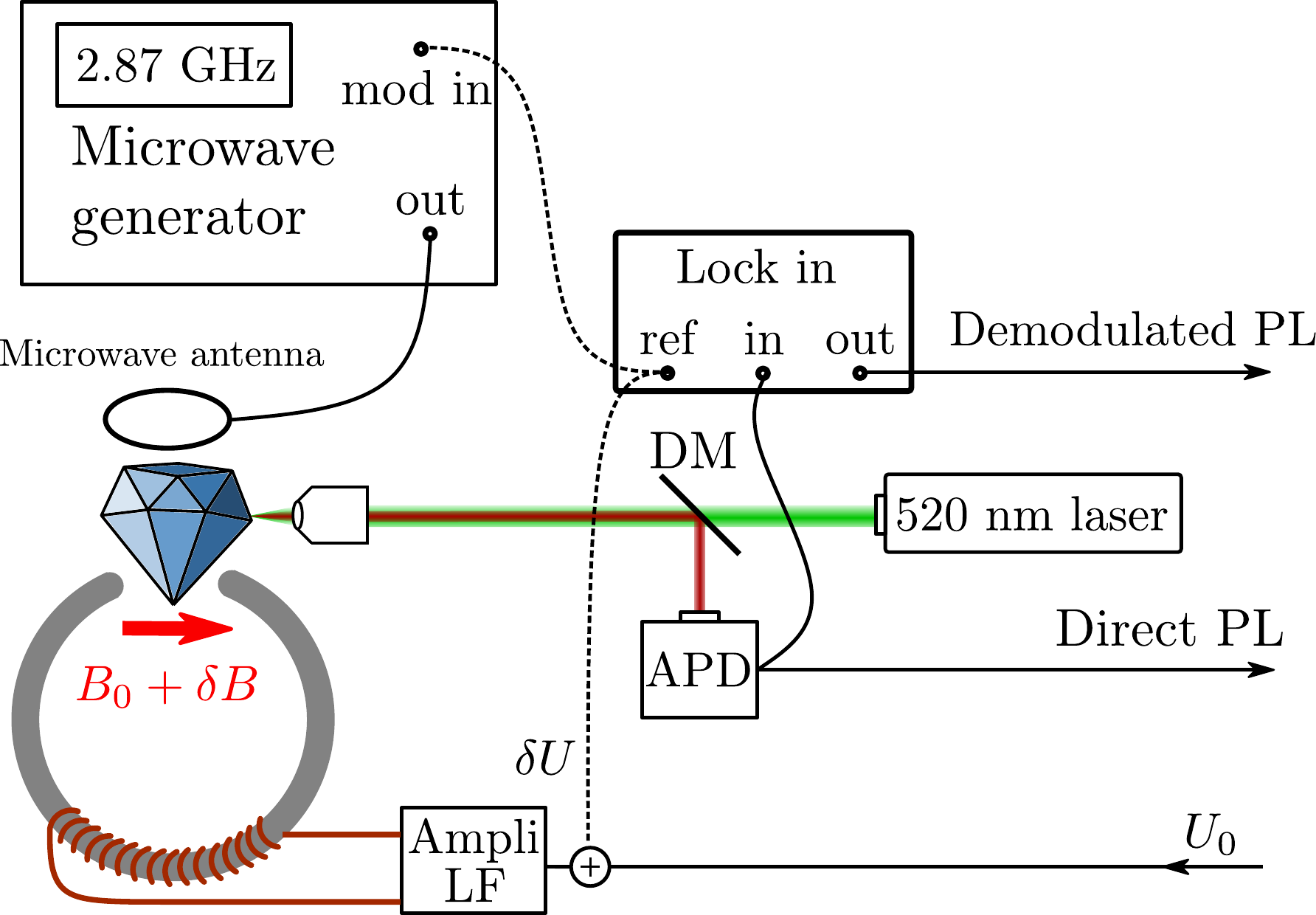}
\caption{Experimental Setup.}
\label{setup}
\end{figure}
Fig. \ref{setup} shows the experimental setup used for all the experiments presented in this article.

The optical polarization and readout of the spins is done by focusing a green laser on the diamond sample with an objective lens (NA=0.65), and collecting the back-scattered red fluorescence from the NV center on an avalanche photo-diode (APD, Thorlabs APD410A). The laser is filtered out using a dichroic mirror and a notch filter. The NV$^0$ fluorescence is filtered using an additional 645 nm long-pass filter.

The laser used here is a Picoquant PDL 800-D with a 520 nm LDH laser head, providing pulses of 40 ps at a rate of 20 MHz, with an average power of $0.5 \sim 5\ \rm mW$. These fast pulses serve no purpose in the actual experiment. This laser was used solely because of its high amplitude stability. The trigger of the pulses is generated externally in order to achieve fast gating of the laser for the $T_1$ measurement. We previously did similar experiment using a continuous 532 nm laser and observed no difference in the spin response. 

The magnetic field is provided by a homemade electromagnet composed of a C-shape iron core and  copper wires. The magnet is mounted on two mechanical rotation stages, allowing a control on the polar and azimuthal angle of the magnetic field within a fraction of a degree and is alimented through a low frequency amplifier (Leybold power function generator 522 63).

The microwave field is generated by a Rhode \& Shwarz SMB 100A and is emitted with a handmade loop antenna. The microwave field is gated by a Mini-circuits ZASWA-2-50DRA+ switch controlled externally and amplified by a Mini-circuits ZHL-5W-422+ amplifier.

A lock-in amplifier (SRS SR830 DSP) is used either to modulate the microwave amplitude for ODMR measurement, or to add an oscillatory magnetic field for the magnetometry protocol. In both case we use a modulation frequency $\sim 1\ \rm kHz$ and demodulate the APD signal.

We did not use magnetic shielding to protect from the earth magnetic field since the splitting due to  $B_{\rm earth}$ is lower than the splitting due to the local electric field for the samples used here : $ \gamma_e B_{\rm earth} \approx 1.5\ \rm MHz < d_\perp E_\perp \approx 4\ \rm MHz$. 

\section{Experimental details}
\subsection{$T_1$ fitting Protocol}
\begin{figure}
\includegraphics[width=0.8\textwidth]{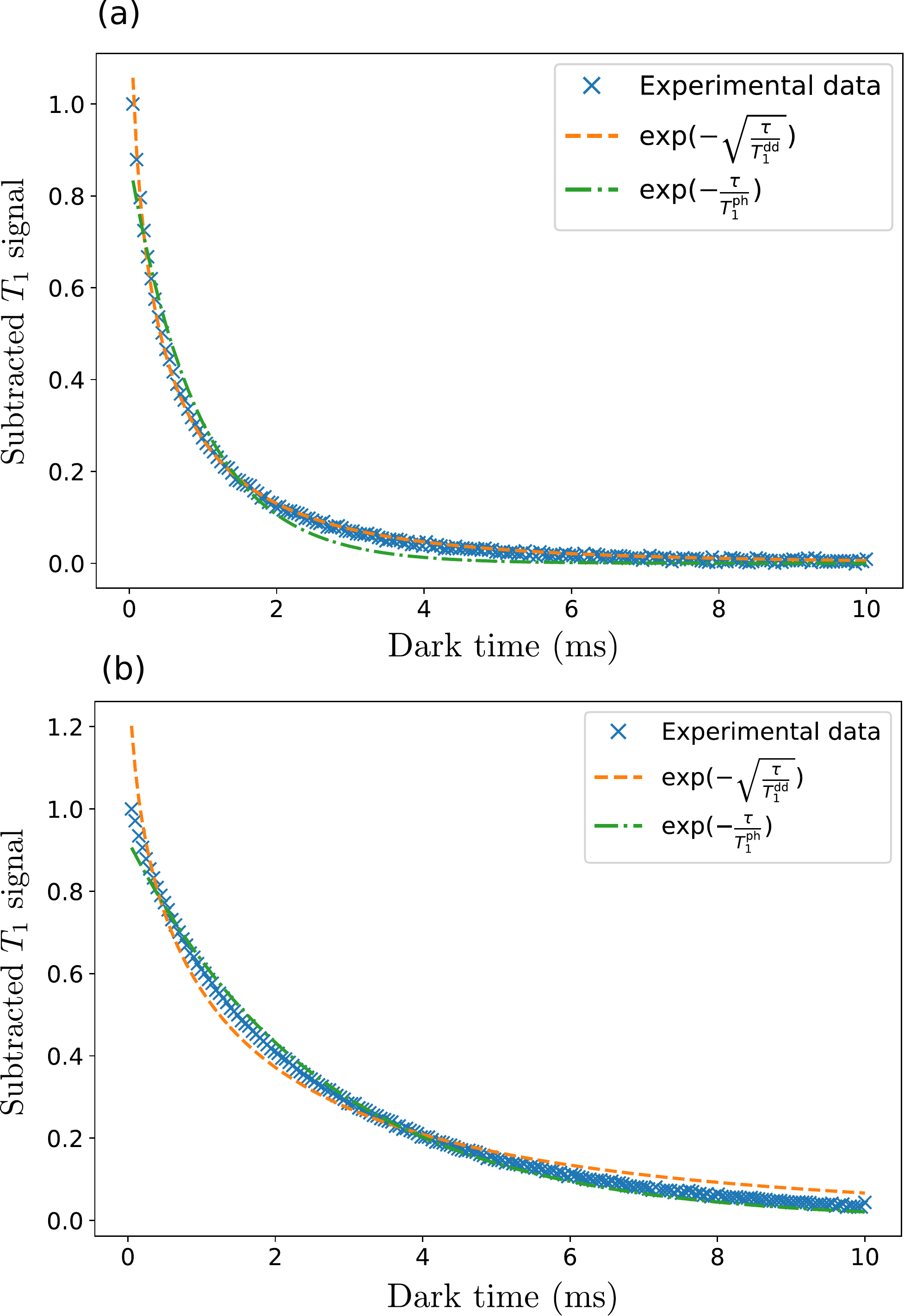}
\caption{$T_1$ measurement with purely exponential and purely stretched exponential fits (a) in zero magnetic field (b) in non-zero magnetic field.}
\label{T1_fits}
\end{figure}
\begin{figure}[h]
\includegraphics[width=0.45\textwidth]{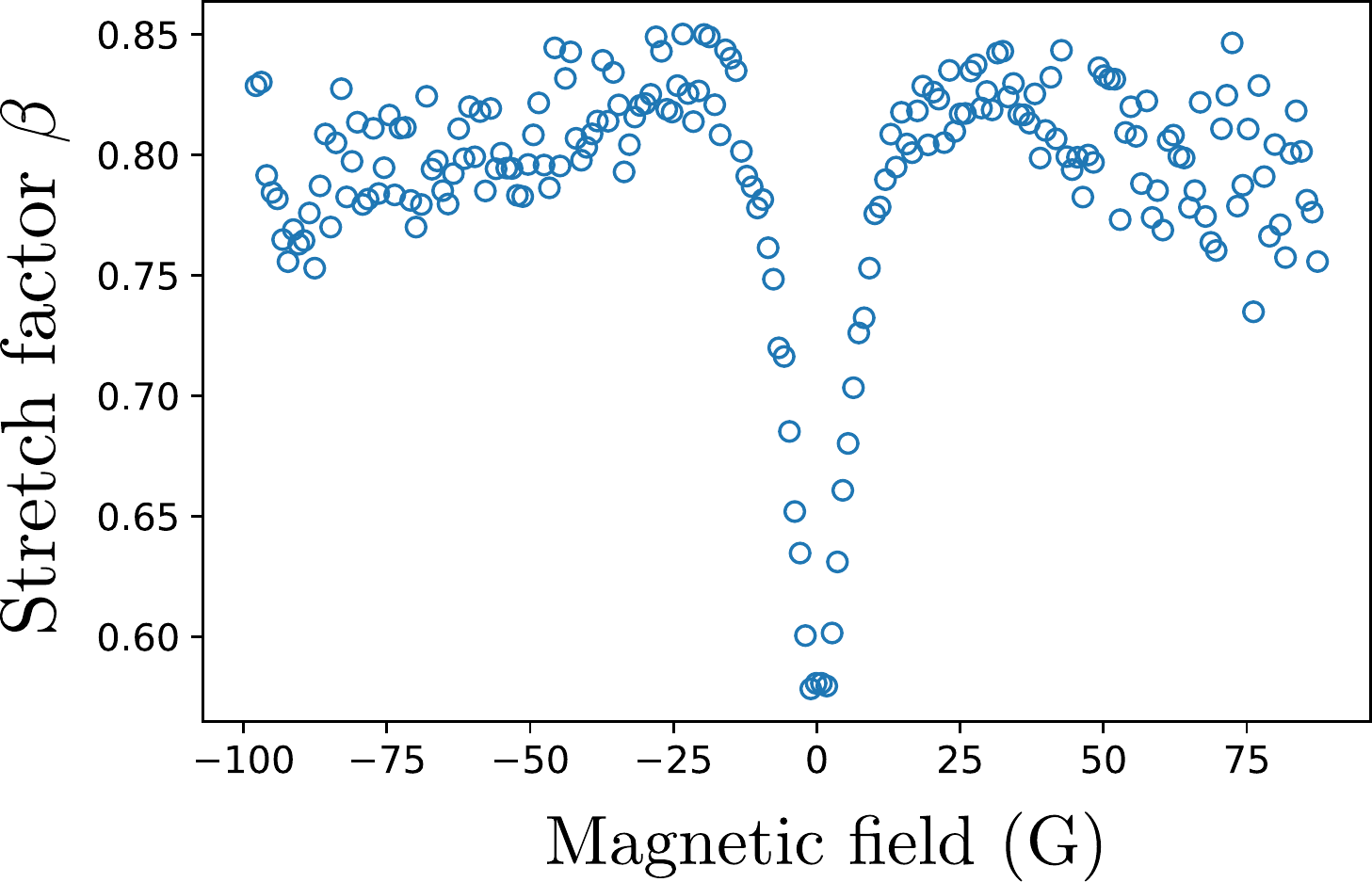}
\caption{Best stretch factor $\beta$ for a $T_1$ fit of the form $f(\tau)=A \exp(-(\frac{\tau}{T_1})^\beta)$ as a function of an arbitrarely oriented magnetic field amplitude.}
\label{alphas}
\end{figure}


Fig. 2-c) in the main text shows the sequence employed for measuring $T_1^{\rm dd}$. 
It consists in a pump-probe measurement where the spins are first polarized in the $\ket{0}$ state by a green laser, and read-out optically after a variable dark time $\tau$. In highly doped samples, this sequence often results in artifacts, mostly due to charge state transfer in the dark \citep{giri_coupled_2018, giri_selective_2019, choi_depolarization_2017}. It is therefore convenient to repeat the sequence with an additional $\pi$ pulse on one of the eight NV spin-resonances right before the spin read-out to prepare the remaining $\ket{0}$ polarization into a darker $\ket{+1}$ or $\ket{-1}$ state \citep{jarmola_temperature-_2012,mrozek_longitudinal_2015,choi_depolarization_2017}. By subtracting the result of the two sequences, we select only the spin-dependent part of the signal, with the added benefit of being able to select a specific class of NV centers.

Fig. \ref{T1_fits} shows the result of a lifetime measurement for the two scenarios presented in main text, namely $B=0$ and $B=50\ \rm G$ in a direction for which every classes are split, on sample HPHT-150-1. In main text, we fitted these results with the expression 
\begin{equation}
S(\tau)=A \exp(-\sqrt{\frac{\tau}{T_1^{\rm dd}}}-\frac{\tau}{T_1^{\rm ph}}),
\label{T_1 formula}
\end{equation}
in order to extract the $T_1$ part associated with the dipole-dipole coupling, as was previously done to study the cross-relaxation between NV and substitutional nitrogen (P1) center \citep{Hall}.

In Fig. \ref{T1_fits}, we present results of $T_1$ measurements when using either purely exponential or purely stretched exponential fits. We can see that for $B=0$, which corresponds to the regime where dipole-dipole relaxation is strong, the experimental data is closely fitted by the stretched exponential profile, whereas for $B=50\ \rm G$, the experimental data follows more closely the exponential profile. In order to describe both regime simultaneously, we have to take both $T_1^{\rm dd}$ and $T_1^{\rm ph}$ into account in our analysis.

Since we are only interested in the stretched exponential part of the lifetime decay and that the phonon-limited exponential decay does not depend on the external magnetic field, we fix the value $T_1^{\rm ph}$ for each sample, in order to reduce the number of free parameters in eq. \ref{T_1 formula}. The value found for all $T_1$ measurements on samples HPHT-150-1 and HPHT-150-2 was $T_1^{\rm ph}=3.62\ \rm ms$.

Another possibility to fit our $T_1$ measurements is the use a varying stretched factor $\beta$ in the fitting function such that :
\begin{equation}
S(\tau)=A \exp (-\left(\frac{\tau}{T_1}\right)^\beta ).
\end{equation}
 Fig. \ref{alphas} shows the optimal $\beta$ parameter as a function of the external magnetic field on sample HPHT-150-1, which confirms that the $T_1$ profile gets closer to a purely stretched exponential in zero magnetic field.

\subsection{Spectral range of the dipole-dipole cross-relaxations}
\label{fluctuator width}
\begin{figure}[h]
\includegraphics[width=0.7\textwidth]{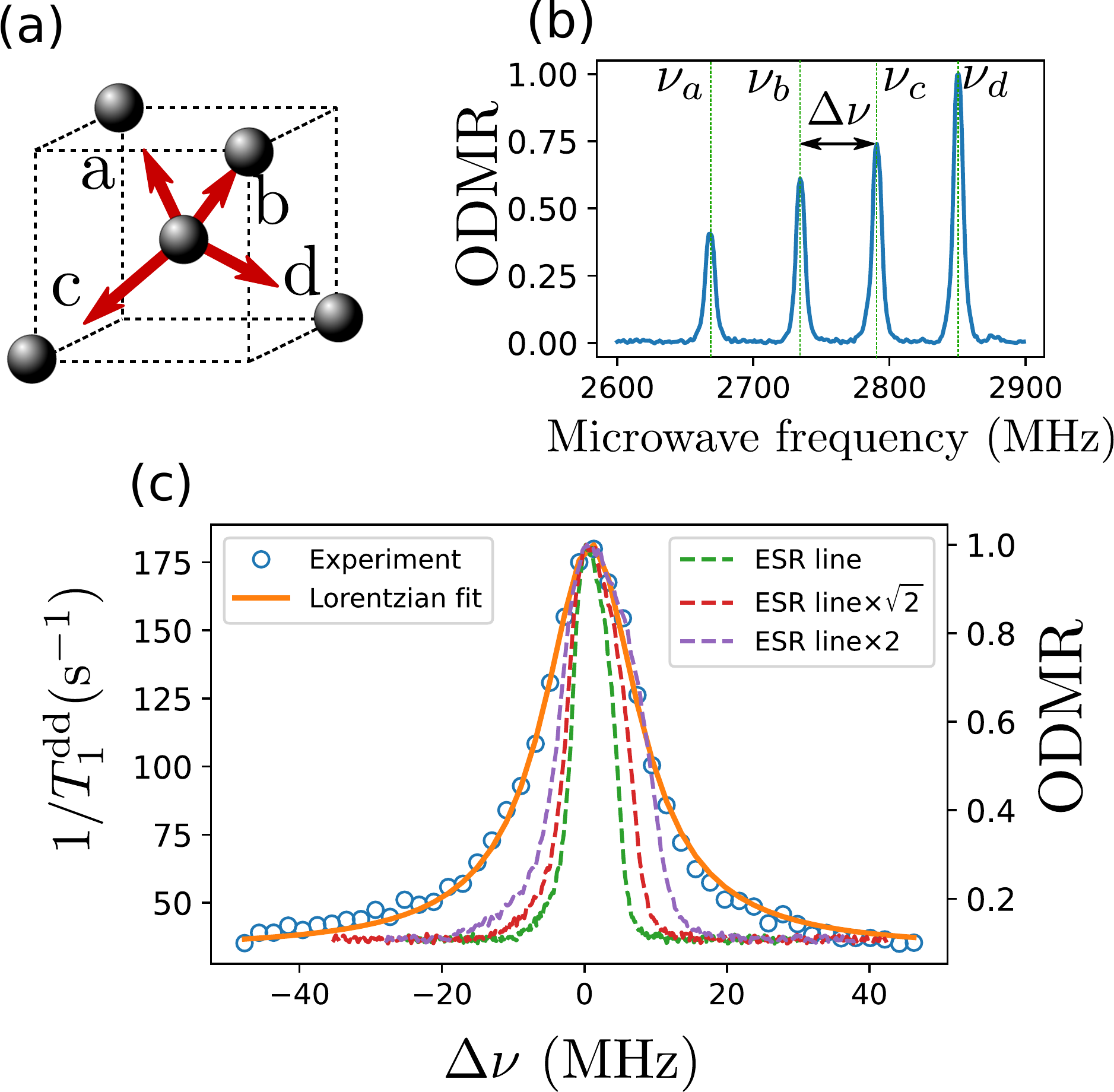}
\caption{Dipole-dipole depolarization for two near-resonant classes. (a) Sketch of the four possible NV orientations (``classes") in a single diamond. (b) ODMR spectrum showing four $\ket{0} \to \ket{-1}$ resonances corresponding to the four spin classes. The detuning $\Delta \nu$ between the classes b and c was controlled by changing the orientation of the external magnetic field. (c) Stretched part of the lifetime decay curve for the spins resonant with $\nu_c$ as a function of the detuning $\Delta \nu$ (blue circles), fitted by a Lorentzian with half width at half maximum 8.04 MHz. Single class ESR line stretched by a factor of 1,$\sqrt{2}$ and 2 are added for comparison.}
\label{largeur_fluct}
\end{figure}

An experimental signature of the fluctuator hypothesis developed in \cite{choi_depolarization_2017} is the dependence of the depolarization rate when two near-resonant classes are brought to resonance : if there are indeed very fast decaying NV centers (fluctuators with lifetime $T_1^f < 100$ ns), then the spectral width of the fluctuator would be greater than $1/T_2^*$.  For simplicity, it is typically assumed that the latter is the same for all spins in the crystal. This means that NV centers would be able to exchange spin quanta (flip-flop) with non-resonant NV fluctuators detuned by $\Delta \nu$ such that $2\pi/T_1^f > \Delta \nu > 2\pi/T_2^*$.

In order to verify this, we measure the spectral overlap between two classes, which in the absence of fluctuator should be on the order of $1/T_2^*$, and compare it to the actual width of the depolarization rate of the spins
as a function of detuning.

Fig. \ref{largeur_fluct} shows the results of such an experiment. We measured the stretched part of the NV's decay rate at each detuning and obtained Fig. \ref{largeur_fluct} -c), which is very well fitted by a Lorentzian of half-width $\sigma=8.04$ MHz. 
We should note that this broadening can not be explained by the direct coherent dipole-dipole interaction induced splitting : for a sample with 3 ppm of NV centers, the average dipole-dipole interaction strength between two nearest NV centers $J_0/r^3 \sim 27\ \rm kHz$ which is several order of magnitude lower than the broadening we observe. 

\begin{figure}[h]
\includegraphics[width=0.7\textwidth]{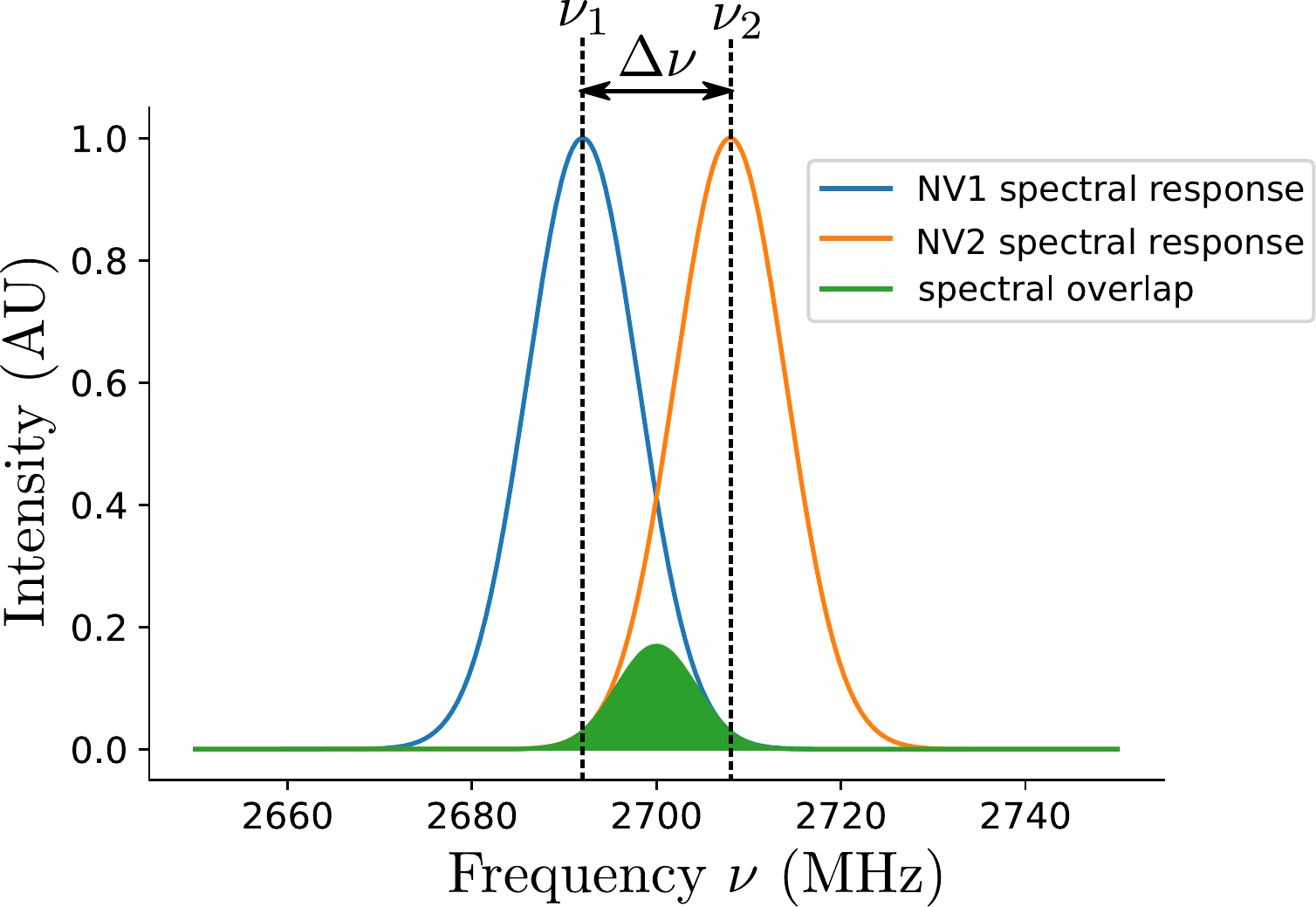}
\caption{Illustration of the spectral overlap for two gaussian spectra.}
\label{overlap}
\end{figure}

We can compare this obtained width to the width of an ODMR line of a single class of NV centers, stretched by a factor of $\sqrt{2}$ and 2 to simulate the spectral overlap of two classes. One should notice that, not only is $1/T_1^{\rm dd}$ significantly larger than the ODMR profile, it also does not have the same shape. 

As illustrated in Fig. \ref{overlap}, we define the spectral overlap $S(\Delta \nu)$ between two spins of spectral response $S_1(\nu)$ and $S_2(\nu)$, centered respectively on the frequencies $\nu_1$ and $\nu_2$ where $\Delta \nu = \nu_2-\nu_1$ as :
\begin{equation}
S(\Delta \nu)=\int S_1(\nu, \nu_1)S_2(\nu, \nu_2) d\nu.
\end{equation}
In order to approximate the spectral overlap in our experiment, we will consider the analytical solution in the Gaussian and Lorentzian case :

\begin{itemize}
\item For two gaussians of standard deviation $\sigma$, the spectral overlap as a function of the detuning $\Delta \nu = \nu_1-\nu_2$ is itself a gaussian of standard deviation $\sigma'=\sqrt{2} \sigma$. :
\begin{align*}
S(\Delta \nu)&\propto \int \exp(-\frac{(\nu-\nu_1)^2}{2\sigma^2})\exp(-\frac{(\nu-\nu_2)^2}{2\sigma^2}) d\nu \\
&\propto\exp(-\frac{(\Delta \nu)^2}{4\sigma^2}).
\end{align*}

\item For two Lorentzian profile with width $\sigma$, the overlap function is itself a Lorentizan with width $\sigma'=2\sigma$ :
\begin{align*}
S(\Delta \nu)&\propto \int \frac{1}{1+ \frac{(\nu-\nu_1)^2}{\sigma^2}}\cdot \frac{1}{1+ \frac{(\nu-\nu_2)^2}{\sigma^2}} d\nu \\
&\propto\frac{1}{1+ \frac{(\Delta \nu)^2}{4\sigma^2}}.
\end{align*}
\end{itemize}

The ODMR lines that we measure are neither Lorentzian nor Gaussian (although they tend to be closer to Gaussians), and can even be asymmetric. Nevertheless, the overlap between two classes can most likely be approximated by a single class ODMR profile stretched by a factor between $\sqrt{2}$ and 2.

\subsection{Lift of the degeneracy between the four classes}

\begin{figure}[h]
\includegraphics[width=0.9\textwidth]{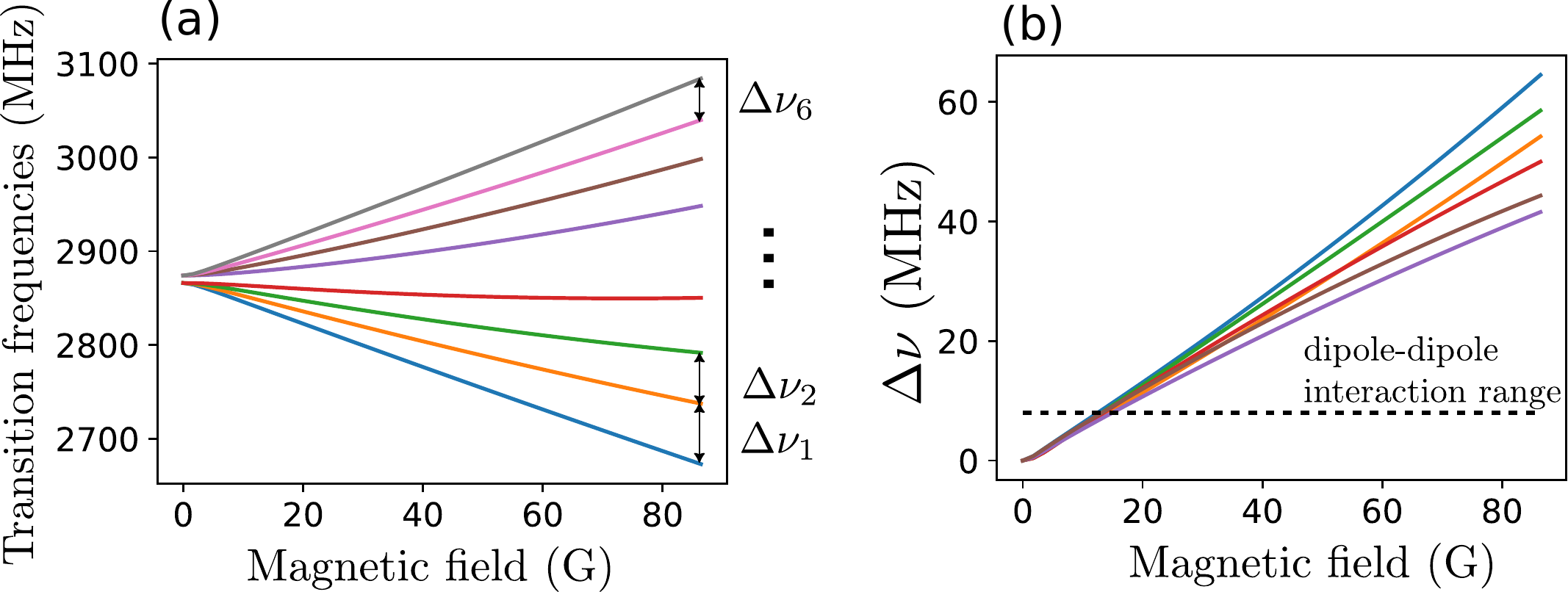}
\caption{(a) : Simulation of the 8 possible transition frequencies (four $\ket{0} \to \ket{-1}$ and four $\ket{0} \to \ket{+1}$ transitions) as a function of the magnetic field amplitude for the orientation of the magnetic field shown in Fig. 3(b) in the main text. (b) Difference in frequency between each pair of closest classes as a function of the magnetic field amplitude. The dotted line y=8.04 MHz corresponds to an estimation of the dipole-dipole interaction frequency range.}
\label{splitting}
\end{figure}
We give here an estimate of the magnetic field required to lift the degeneracy of the four classes in the case of Fig. 3(b) in the main text. Fig. \ref{splitting} shows the simulated energies for the 4 $\ket{0} \to \ket{-1}$ and the 4 $\ket{0} \to \ket{+1}$ transitions as a function of the magnetic field, and the difference in energy between the closest pairs of transitions. We can see that the energy detuning between the classes $\Delta \nu$ crosses the value of the dipole-dipole interaction range for magnetic field values $\norm{B} = 12 \sim 15\ \rm G$, which are close to the half-width of the zero-field feature in Fig. 3(b)(ii) and (iii) in the main text.

\subsection{Effect of laser polarization}
Previous studies \cite{anishchik2015low, filimonenko2020weak} reported the presence of a photoluminescence dip in zero magnetic field which was heavily dependent on the laser polarization angle with respect to the diamond axes and the magnetic field. 

\begin{figure}[h]
\includegraphics[width=0.9\textwidth]{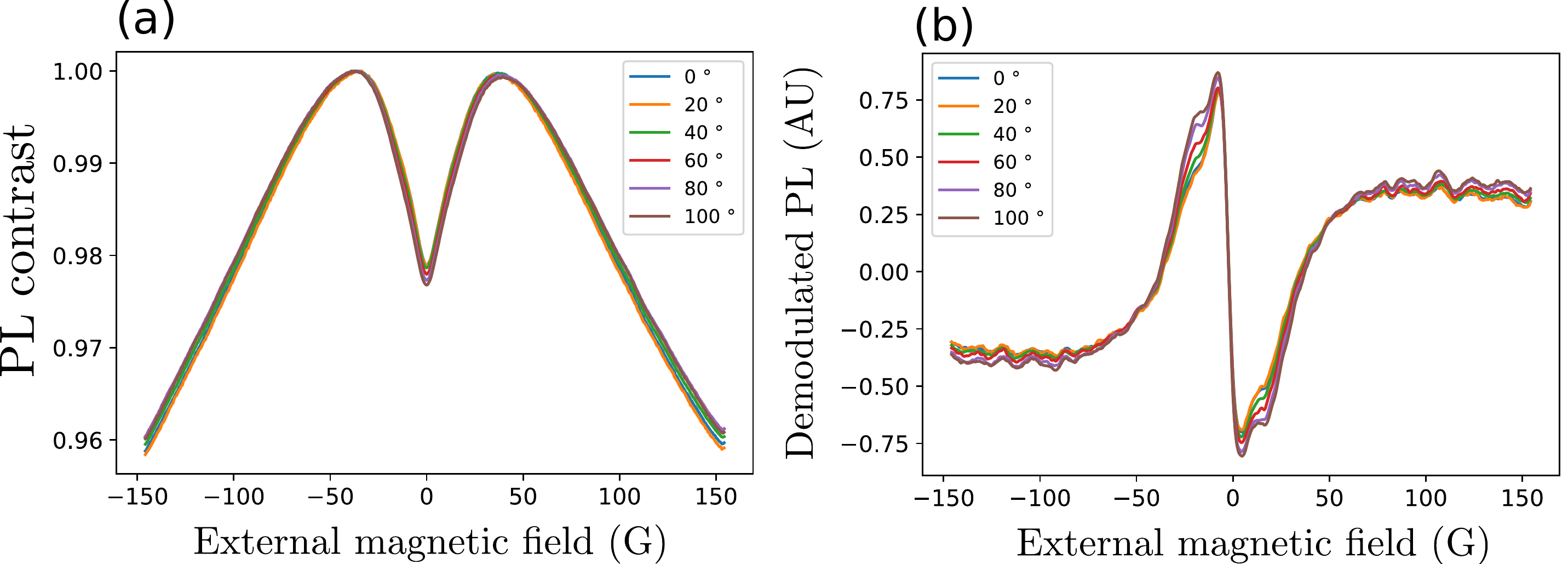}
\caption{Effect of the polarization of the incident laser. (a) Photoluminescence of sample HPHT-150-3 as a function of arbitrarely oriented magnetic field amplitude for various polarization angle. (b) Demodulated PL in the same conditions.}
\label{Pola}
\end{figure}

We did not observe a strong dependence on the laser polarization with the samples used in this study. Fig.\ref{Pola} shows the photoluminescence from one of our sample as a function of the magnetic field, either directly or with a modulation of the magnetic field, for 6 polarization angles and saw no major difference. 

In particular, we did not observe the apparition of an anti-line inside the main dip, unlike what was observed in the two previously cited work. We expect that the main reason behind the different behaviors is that we seem to observe far greater spin depolarization in zero-field, which we attribute to dipole-dipole coupling. This effect could hide smaller effects such as the one related to the laser polarization.
\subsection{Alignment of the magnetic field along [100] axis}
\begin{figure}[h]
\centering
\includegraphics[width=0.6\textwidth]{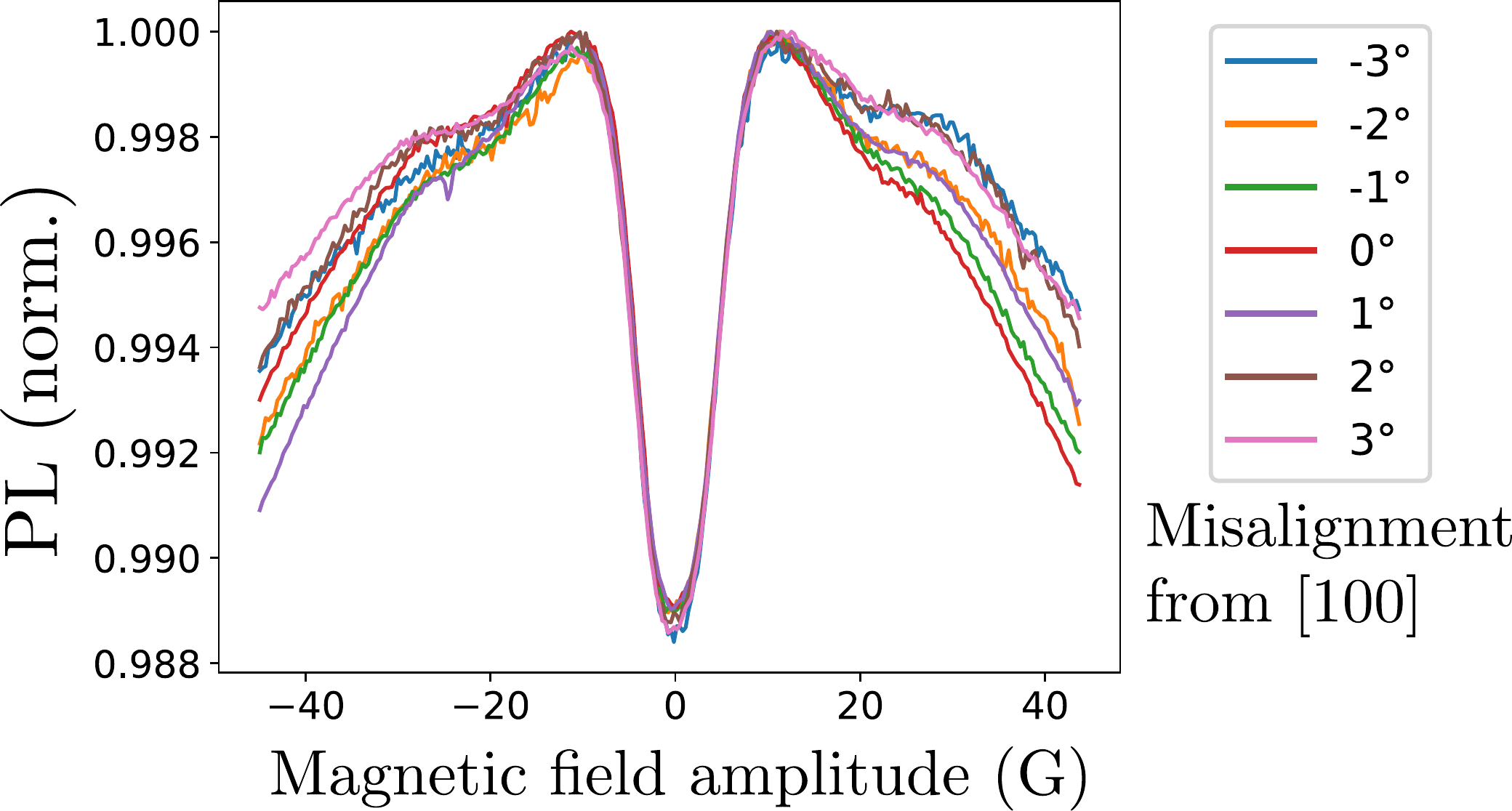}
\caption{PL of sample CVD-PPM as a function of the external magnetic field amplitude for different magnetic field misalignments with respect to the [100] axis.}
\label{Alignment}
\end{figure}
We claim that the decrease in PL at low magnetic field in Fig. 3 (c)(ii) in the main text is due to the local electric field and the double flip processes. One other possibility would be a misalignment of the magnetic field with respect to the [100] axis as it is scanned. The four classes could for instance be truly resonant only with zero external magnetic field and not when the B field is large.

Fig. \ref{Alignment} shows the change in PL with respect to the external magnetic field amplitude for various misalignments of the magnetic field direction compared to the [100] diamond crystalline axis of the sample CVD-2. We can see that the central drop in PL is not critically affected by the misalignment of the magnetic field, and we are therefore confident that the drop in PL we observe when $B\parallel [100]$ is not an effect of misalignment.

The initial alignment of the magnetic field was controlled by applying a field of $\sim 50\ \rm G$ and monitoring the overlap of the four classes with ODMR spectra (maximizing the PL drop is also an effective method). We estimate the initial alignment precision to be $\pm 1^\circ$. The misalignment was then introduced by rotating the electromagnet along an arbitrary axis. The slight asymmetry between the positive and negative value of the magnetic field could come from the earth magnetic field or from hysteresis in the magnetic core of our electromagnet.

\subsection{Emulation of the electric field with transverse magnetic field}
\begin{figure}
\includegraphics[width=0.9\textwidth]{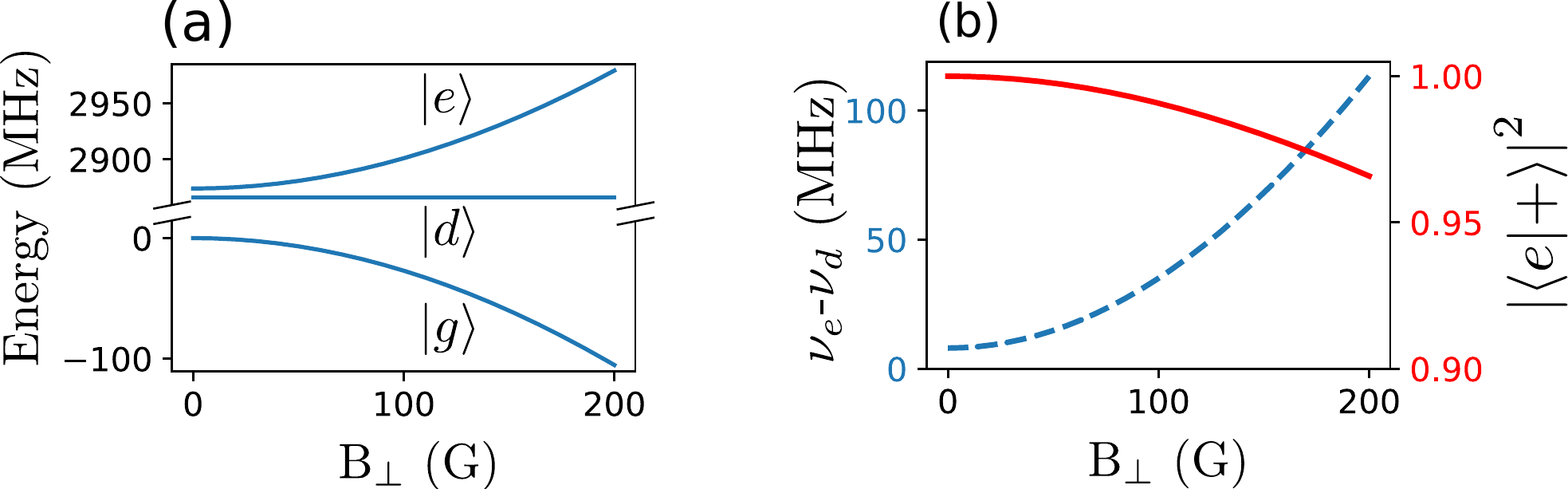}
\caption{(a) Simulated energy level of the three eigenstates $\ket{g},\ket{d}$ and $\ket{e}$ for an NV center under purely transverse magnetic field. The spin Hamiltonian is given in eq. \ref{NV Hamiltonian} and a value of $d_\perp E_\perp = 4\ \rm MHz$ was chosen (b) Blue dashed curve : frequency splitting between the $\ket{d}$ and $\ket{e}$ states as a function of the transverse magnetic field, red plain line : matching factor $\abs{\bra{e}\ket{+}}^2$ between the $\ket{e}$ and $\ket{+}$ states.}
\label{theory_transverse_field}
\end{figure}
In the main text we use the property that, under purely transverse magnetic field, the eigenstates of the single spin Hamiltonian are close to the $\{ \ket{0},\ket{+},\ket{-} \}$ basis, described in the main text and in sec. \ref{sec Hamiltonian} of these supplementary. We will discuss here the validity of this claim.

We will call the true eigenstates of the single spin Hamiltonian under purely transverse magnetic field $\{ \ket{g},\ket{d},\ket{e} \}$ in ascending order of energy. The $\ket{d}$ state is strictly equal to the $\ket{-}$ state. The $\ket{g}$ and $\ket{e}$ states can be written as a mixture of the $\ket{0}$ and $\ket{+}$ states, and their exact expression depends on the field value. As $B_{\perp}$ approaches 0 however, $\ket{g}$ and $\ket{e}$ tend toward $\ket{0}$ and $\ket{+}$ respectively.

Fig. \ref{theory_transverse_field} (a) shows a simulation of the energy levels of the three eigenstates $\ket{g}$, $\ket{d}$ and $\ket{e}$ as a function of the transverse magnetic field amplitude. Of importance for the results in main text is the splitting $\Delta \nu$ between the $\ket{e}$ and $\ket{d}$ states, which is equal to the difference in the transition frequencies $\ket{g} \to \ket{d}$ and $\ket{g} \to \ket{e}$ (reported experimentally as $\nu_+$ and $\nu_-$ in main text). 

This value $\Delta \nu$ is reported on Fig. \ref{theory_transverse_field} (b) along with a matching factor $\abs{\bra{e}\ket{+}}^2$. This factor is always equal to $\abs{\bra{g}\ket{0}}^2$ since $\bra{d}\ket{-}=1$. We can see that indeed, for $B_{\perp}<150\ \rm G$, $\abs{\bra{e}\ket{+}}^2>0.98$. Crucially, for $B_{\perp}=150\ \rm G$, the splitting $\Delta \nu$ is equal to $\approx 70\ \rm MHz$, almost an order of magnitude greater than the dipole-dipole interaction range computed on sec. \ref{fluctuator width}. This means that we can effectively quench the double-flips by splitting the $\ket{e}$ and $\ket{d}$ states, while still maintaining eigenstates almost equal to $\ket{0}$, $\ket{+}$ and $\ket{-}$.

\section{Extension of the fluctuator model to low magnetic fields}

\subsection{Results}

Here are the results of the fluctuator model developed in \cite{choi_depolarization_2017} that we extended to include double flip as well as local electric fields. The following sections will detail the calculations that led to these results. Let us first summarize our theoretical findings :

\begin{itemize}
\item The dipole-induced spin decay is lowest for a spectrally isolated class of NV centers when the transverse magnetic field is negligible. In this case the dipole-dipole relaxation is limited by flip-flop processes between spins of the same class. We will call the corresponding stretched lifetime $T_1^{\rm dd}\equiv T_0$.
\item For a single class, still spectrally isolated   from the three other classes, but this time dominated by the electric field or transverse magnetic field (as long as $\gamma_e B_\perp \ll D$), the change in the eigenstates of the single particle Hamiltonian results in an increase of the average flip-flop rate. This leads to a new theoretical stretched lifetime $T_1^{\rm dd}= 4\, T_0$.
\item When a magnetic field is applied along the [100] axis (for $5\ {\rm G} < \abs{\bm B} \ll D/\gamma_e$), all four classes are resonant and dominated by the longitudinal part of the magnetic field. This results, in increase in the average flip-flop rate compared to the isolated class case, and the predicted stretched exponential lifetime is $T_1^{\rm dd}\approx 42.8\, T_0$.
\item In zero external magnetic field, all four classes are also resonant but the single particle Hamiltonian of each spin is dominated by a random local electric field. The averaging of flip-flop in this case yields a stretched lifetime $T_1^{\rm dd}\approx 51.4\, T_0$, which is about 20\% higher than the predicted lifetime for $\bm B \parallel$ [100].
\end{itemize}

The effect of the double flip processes was not studied quantitatively since the process is never fully resonant and is therefore highly dependent on the initial splitting of the $\ket{+}$ and $\ket{-} $ states. Nevertheless, the presence of double-flip as an additional relaxation channel can only increase the dipole-induced spin decay rate. 

\subsection{Summary of the NV-fluctuator model}
Here are the main hypotheses and conclusion of the NV-fluctuator model developed in \cite{choi_depolarization_2017} :

The NV$^-$ centers in the crystal are divided between two categories :``normal" NV centers (simply called NV) which, in the absence of dipole-dipole coupling would have a phonon-limited $T_1$ ($T_1^{NV}\sim \rm ms$), and fluctuators who are NV centers with an additional, fast, depolarization mechanism, such that their lifetime $T_1^f < 100\ \rm ns$. Having such a short lifetime, the fluctuators are almost unpolarized by the green laser, making them invisible in standard optical NV measurement protocol ($T_1$, ODMR, etc.). 

Assuming a homogeneous distribution of the fluctuators in the bulk of the crystal, the authors of \cite{choi_depolarization_2017} show that the fluctuators create an additionnal decay channel for the NV population, through dipole-dipole interaction, characterized by a decay rate $\gamma$ which follows the probability distribution :
\begin{equation}
\rho(\gamma)=\frac{e^{-1/(4\gamma T)}}{\sqrt{4\pi \gamma^3 T}},
\end{equation}
where the timescale $T$ is defined as
\begin{equation}
\frac{1}{T}=\left(\frac{4\pi n_fJ_0\bar \eta}{3}\right)^2 \frac{\pi}{\gamma_f},
\label{eq 1/T}
\end{equation}
where $n_f$ is the fluctuator density in the crystal in nm$^{-3}$, $J_0=52\ \rm{MHz}\cdot\rm{nm}^3$ is the characteristic dipole-dipole strength between two spins, $\gamma_f$ is the fluctuator intrinsic decay rate and $\bar \eta$ is a dimension-less number which characterizes the average dipole-dipole interaction between the NV centers and the fluctuators (resonance conditions, relative orientations etc. Further details are given below).

The polarization dynamics of the averaged ensemble of NV centers then follows :
\begin{equation}
P(t)=\int_0^\infty \rho(\gamma)\, e^{-\gamma t}d\gamma= e^{-\sqrt{t/T}}.
\end{equation}
Which corresponds to the stretched-exponential part of the lifetime measurement.
\subsection{Dipole-Dipole Hamiltonian between two NV$^-$ centers}
Let us consider the hamiltonian of an NV center interacting with a fluctuator :
\begin{equation}
\mathcal{H}_{\rm tot}=\mathcal{H}_{1}+\mathcal{H}_2+\mathcal{H}_{\rm dd}.
\end{equation}
$\mathcal{H}_{1}$ and $\mathcal{H}_{2}$ are the single particle Hamiltonian of the NV and fluctuator, described by the equation (\ref{NV Hamiltonian}), and $\mathcal{H}_{\rm dd}$ is the dipole-dipole interaction Hamiltonian between the two spins. 

We will consider the spin operators $\bm{S}_1$ and
$\bm{S}_2$ in the NV basis : the $z$ orientation of the $S_z$ operator is chosen along the NV axis of each spin. In the presence of a longitudinal magnetic field, the sense of $\hat{z}$ will be chosen such that $\bm{B}\cdot \hat{z} >0$, and the $\hat{x}$ direction is chosen along the transverse magnetic or electric field, both of which dominate the single spin Hamiltonian. These choices mean that there will be two distinct Cartesian basis $\{\hat{x}_1,\hat{y}_1,\hat{z}_1\}$ and $\{\hat{x}_2,\hat{y}_2,\hat{z}_2\}$ describing the two spins.

The dipole-dipole Hamiltonian can then be decomposed as :
\begin{align}
-\frac{\mathcal{H}_{\rm dd}}{J_0/r^3}&= 3\left({\bm S}_1 \cdot \hat{u} \right)\left({\bm S}_2 \cdot \hat{u} \right) - {\bm S}_1 \cdot {\bm S}_2  \\
&=\left[3(\hat{u}\cdot\hat{x}_1)(\hat{u}\cdot\hat{x}_2) -\hat{x}_1\cdot\hat{x}_2\right] S_x^1S_x^2 \\
&+\left[3(\hat{u}\cdot\hat{y}_1)(\hat{u}\cdot\hat{y}_2) -\hat{y}_1\cdot\hat{y}_2\right] S_y^1S_y^2 \\
&+\left[3(\hat{u}\cdot\hat{x}_1)(\hat{u}\cdot\hat{y}_2) -\hat{x}_1\cdot\hat{y}_2\right] S_x^1S_y^2 \\
&+\left[3(\hat{u}\cdot\hat{y}_1)(\hat{u}\cdot\hat{x}_2) -\hat{y}_1\cdot\hat{x}_2\right] S_y^1S_x^2 \\
&+\left[3(\hat{u}\cdot\hat{z}_1)(\hat{u}\cdot\hat{z}_2) -\hat{z}_1\cdot\hat{z}_2\right] S_z^1S_z^2, \\
&+\mathcal{H}_{\rm other},
\end{align}
where $J_0=\frac{\mu_0 \gamma_e^2 \hbar^2}{4 \pi} = (2 \pi) 52\ \rm MHz \cdot \rm{nm}^3$, $\bm r= r\bm{\hat u}$ is the distance between the two spins and  $\mathcal{H}_{\rm other}$ contains terms of the form $S_x^iS_z^j$ and $S_y^iS_z^j$ which couple states far from resonance and will be neglected here.

\subsection{Flip-flops in the magnetic basis $\{ \ket{0},\ket{+1},\ket{-1} \} $}
We will first consider the case, treated in \cite{choi_depolarization_2017}, where the longitudinal magnetic field is strong enough that the single spin Hamiltonian eigenstates of both spins are close to $\{ \ket{0},\ket{+1},\ket{-1} \} $ (since we are interested in near-resonant spins, this means that both spins see roughly the same longitudinal and transverse magnetic field).

In this scenario, only flip-flop terms (that is $\mel{\pm 1,0}{\mathcal{H}_{\rm dd}}{0,\pm 1}$) can couple two resonant two-spins states. Following the notation in \cite{choi_depolarization_2017}, we introduce the dimensionless factor $\eta$ defined as : 
\begin{equation}
\eta^2=\frac{1}{3} \abs{\mel{\pm 1,0}{\frac{\mathcal{H}_{\rm dd}}{J_0/r^3}}{0,\pm 1}}^2  \frac{4\gamma_f^2}{(\omega_f - \omega_{NV})^2+4\gamma_f^2},
\end{equation}
where $\gamma_f$ is the fluctuator lifetime, $\omega_f/2\pi$ is the transition frequency of the fluctuator and $\omega_{NV}/2\pi$ the NV transition frequency. 

This numerical factor is linked to the previously mentioned $\bar \eta$ by averaging over every possible orientation and angular position of the fluctuators. Assuming that the fluctuators are evenly distributed among all four classes of NVs and decomposing the $\bm r$ vector in the spherical basis ($r, \theta, \phi$), we can write $\bar \eta$ as :

\begin{equation}
\label{eq. eta bar}
\bar \eta =\frac{1}{4} \sum_{i=1}^4 \int_\theta \int_\phi  \int_{\omega_f} \int_{\omega_{NV}}   \abs{\eta(\bm u,i,\omega_{NV},\omega_f)}\, d\Omega\, \rho(\omega_{NV}) d\omega_{NV}\, \rho^i(\omega_f) d\omega_f ,
\end{equation}

where $d\Omega=\sin \theta d\theta d\phi$, $\rho(\omega_{NV})$ is the distribution of angular frequencies of the probed NV centers, and  $\rho^i(\omega_f)$ is the distribution of angular frequencies  for the fluctuators of class $i$.

While the distributions $\rho(\omega_{NV})$ and $\rho^i(\omega_f)$ can be approximated using the data presented in Fig. \ref{largeur_fluct}, we will assume that these distributions are independent of the external magnetic field, and since we are interested in comparing the same sample for various values of the magnetic field, we will simplify eq.(\ref{eq. eta bar}) by setting $\frac{4\gamma_f^2}{(\omega_f - \omega_{NV})^2+4\gamma_f^2}$ to 1 if the classes of NV centers and fluctuators are resonant (or if they are from the same class), and to 0 otherwise.

\begin{figure}
\includegraphics[width=0.7\textwidth]{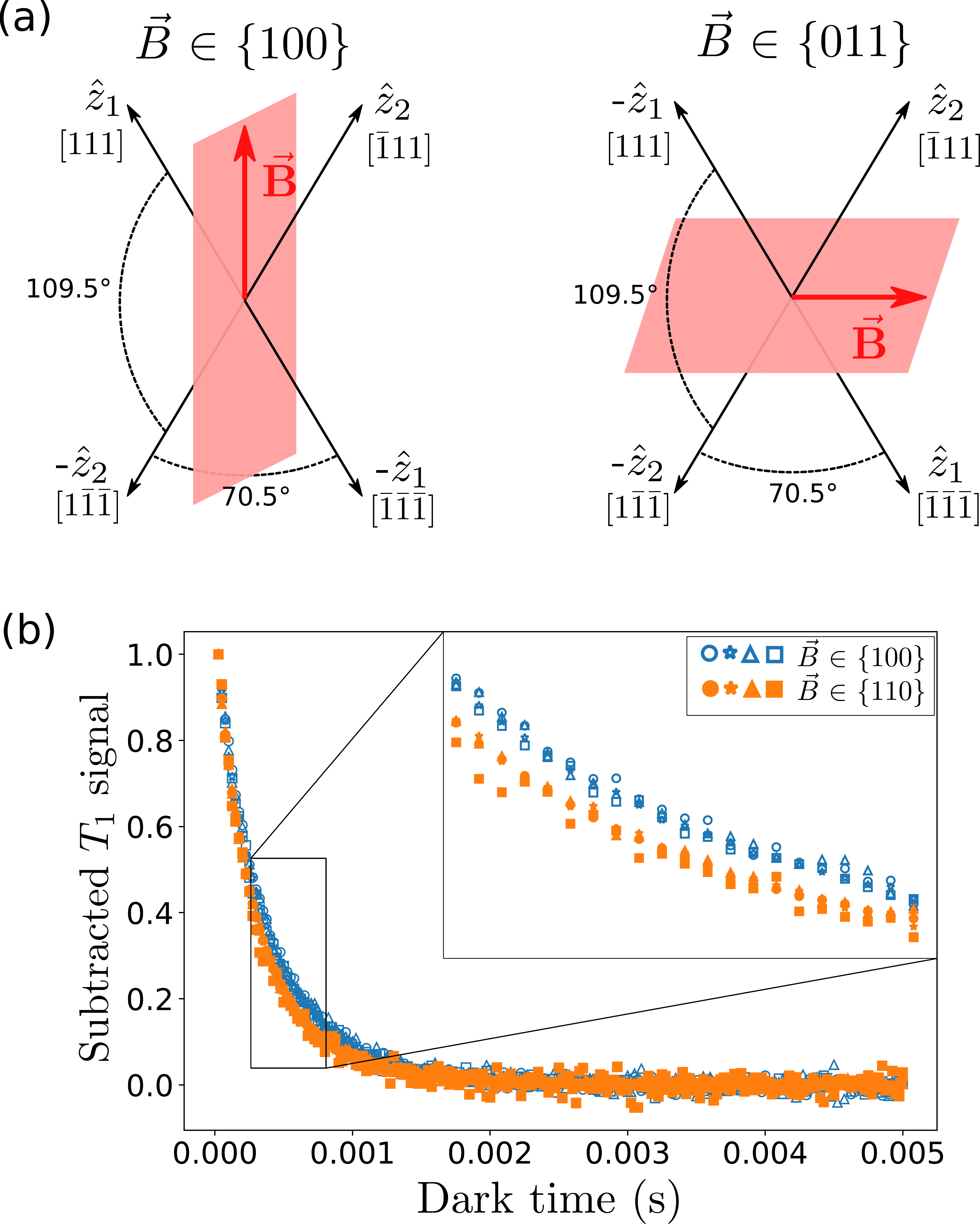}
\caption{(a) Geometrical representation of two NV classes and the two possible planes where the magnetic field has the same projection on both classes. The positive $\hat z_i$ direction is chosen so that $\bm{B}\cdot \hat{z}_i >0$. (b) $T_1$ measurement on a two-class resonance following the protocol described in main text. 8 different magnetic field were employed on the same sample : 4 times in a \{100\} plane (blue unfilled symbols) and 4 times in a \{110\} plane (orange filled symbols). Inset : zoom-in on the 0.3-0.8 ms region.}
\label{121 VS 22 fig}
\end{figure}

With these hypotheses, we find that there are 3 possible scenarios where the NV and fluctuator are resonant :
\begin{itemize}
\item The NV and fluctuator are from the same class, so that the angle $\widehat{z_1 z_2}=0$ : this is always true, regardless of the magnetic field angle and amplitude. We can analytically compute $\bar \eta$ in this case and find : $$ \bar {\eta}_{same}=\frac{1}{4}  \int_\theta \int_\phi \abs{\eta(\bm u,\widehat{z_1 z_2}=0,\omega_{NV}=\omega_f)}\, d\Omega= \frac{1}{4} \cdot \sqrt{\frac{1}{3}} \cdot \frac{2}{3 \sqrt{3}} \approx 5.55\cdot10^{-2}.$$
\item The angle $\widehat{z_1 z_2}=\rm{arccos}(\frac{1}{3})\approx 70.5^\circ$. This is the case when the magnetic field lies in the \{100\} crystalline planes family. In this case, numerical simulations yield $\bar \eta_{\rm close} \approx \frac{1}{4} \cdot \sqrt{\frac{1}{3}} \cdot 0.6507 \approx 9.39\cdot10^{-2}$
\item The angle $\widehat{z_1 z_2}=\rm{arccos}(-\frac{1}{3})\approx 109.5^\circ$. This happens when the magnetic field lies in the \{110\} or \{$1\bar{1}0$\} crystalline planes family. In this case, numerically simulations give $\bar \eta_{\rm far} \approx \frac{1}{4} \cdot \sqrt{\frac{1}{3}} \cdot 0.8328 \approx 1.20\cdot10^{-1}$. This last case was not taken into account in \cite{choi_depolarization_2017}.
\end{itemize}

Fig. \ref{121 VS 22 fig}(a) shows a graphical representation of the difference between the two last scenarios : due to the $\bm{B}\cdot \hat{z} >0$ condition, the same two classes of NV centers can have a $\widehat{z_1 z_2}$ angle equal to $70.5^\circ$ or $109.5^\circ$ depending on the external magnetic field. 

These computed values can be tested experimentally. Fig. \ref{121 VS 22 fig}(b) shows the subtracted $T_1$ signal measured on the same sample HPHT-15-2 for 8 different values of the magnetic field, each time on a two-class resonance. In 4 cases, the magnetic field was in a \{100\} plane, and in the 4 other cases it was in a \{110\} plane. We can see that the measured lifetime is always smaller on the \{110\} case, which corresponds the the greater $\bar \eta$ factor computed previously.

\begin{figure}
\includegraphics[width=0.7\textwidth]{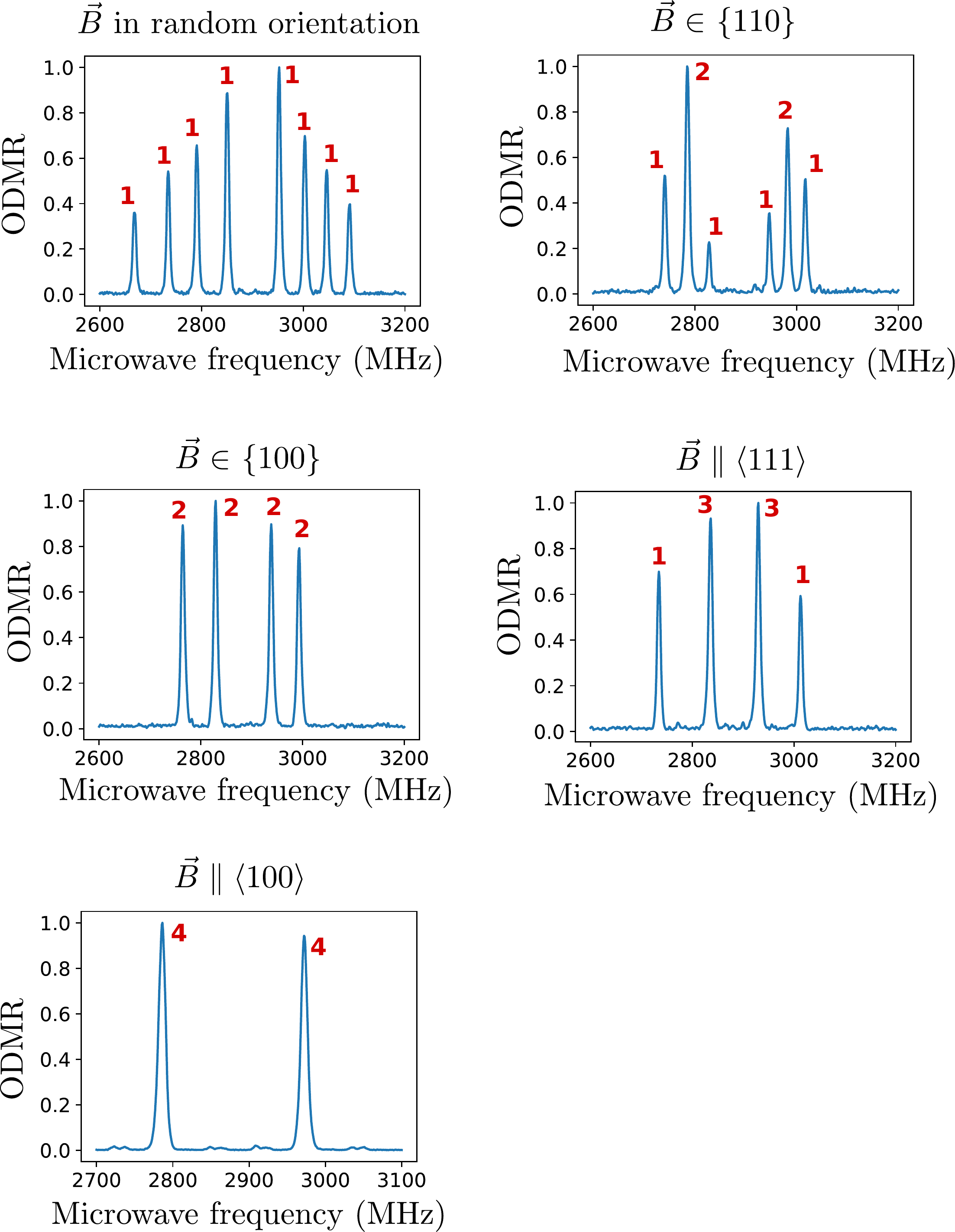}
\caption{ODMR spectra for different orientations of the magnetic field. Red numbers represent the number of degenerate classes for each line of the spectrum.}
\label{Various ODMR}
\end{figure}

Here is a list of the computed values of $\bar \eta^2$, which according to eq.(\ref{eq 1/T}) is proportional to the dipole induced decay rate, for various magnetic field orientation. An ODMR spectrum for each of these situations is present in Fig. \ref{Various ODMR}.
\begin{itemize}
\item Random orientation / no class degeneracy : for all four classes, $\bar \eta = \bar {\eta}_{\rm same}$ and $\bar \eta^2=3.08 \cdot 10^{-3} \equiv \bar \eta_0^2$.
\item $\bm{B} \in$ \{110\} : For the two non-resonant classes : $\bar \eta = \bar {\eta}_{\rm same}$ and $\bar \eta^2= \bar \eta_0^2$.  For the two resonant classes : $\bar \eta = \bar {\eta}_{\rm same} + \bar {\eta}_{\rm far}$ and $\bar \eta^2\approx\, 10.0 \bar \eta_0^2$.
\item $\bm{B} \in$ \{100\} : For each pair of two-classes resonance: $\bar \eta = \bar {\eta}_{\rm same}+\bar {\eta}_{\rm close}$ and $\bar \eta^2\approx 7.24\, \bar \eta_0^2$. 
\item $\bm{B} \parallel \langle 111 \rangle$ : For the non-resonant class (parallel to $\bm B$) : $\bar \eta = \bar {\rm \eta}_{same}$ and $\bar \eta^2= \bar \eta_0^2$. For the triply resonant classes: $\bar \eta = \bar {\eta}_{\rm same}+2\bar {\eta}_{\rm far}$ and $\bar \eta^2\approx 28.4\, \bar \eta_0^2$. 
\item $\bm{B} \parallel \langle 100 \rangle$ : For the quadruple resonance : $\bar \eta = \bar {\eta}_{\rm same}+2\bar {\eta}_{\rm close} + \bar {\eta}_{\rm far}$  and $\bar \eta^2\approx 42.8\, \bar \eta_0^2$.
\end{itemize}

The measured increase in the decay rate is generally smaller than the predicted one : we measured an increase by a factor $\sim 4$ for a two-class degeneracy instead of the predicted factor of $10$, and an increase by a factor $\sim 16$ for a four-class degeneracy instead of the predicted factor of $42.8$. \cite{choi_depolarization_2017} measured an increase by a factor $\sim 4$ on a two-class degeneracy instead of the predicted factor of $7.24$ (when $\bm{B} \in$ \{100\}) or $10.0$ (when $\bm{B} \in$ \{110\}).

\subsection{Flip-flops in the non-magnetic basis $\{ \ket{0},\ket{+},\ket{-} \} $}

As described in sec. \ref{sec Hamiltonian}, when the single spin Hamiltonian of the NV center is dominated by the electric field or by the transverse magnetic field in the $\gamma_e B_\perp \ll D$ regime, the eigenbasis of the NV Hamiltonian is close to $\{\ket{0},\ket{+}=\frac{\ket{+1}+\ket{-1}}{\sqrt{2}},\ket{-}=\frac{\ket{+1}-\ket{-1}}{\sqrt{2}} \} $. 
We wish to include the effect of the electric field or transverse magnetic field on the depolarization, so we write the dipole-dipole Hamiltonian in this new basis for both spins.

Since the flip-flop processes are now of the form $\ket{0,\pm}\bra{\pm,0}$, we redefine the $\eta$ factor as :
\begin{equation}
\eta^2=\frac{1}{3} \abs{\mel{\pm ,0}{\frac{\mathcal{H}_{\rm dd}}{J_0/r^3}}{0,\pm }}^2  \frac{4\gamma_f^2}{(\omega_f - \omega_{NV})^2+4\gamma_f^2}.
\end{equation}

In this new basis, the spin operators are written :

\begin{equation}
  S_x = \begin{pmatrix}
  0&1&0 \\
  1&0&0 \\
  0&0&0
  \end{pmatrix}
  \begin{matrix}
  \bra{-} \\
  \bra{0} \\
  \bra{+}
  \end{matrix}  
  , \quad 
 S_y = \begin{pmatrix}
  0&0&0 \\
  0&0&1 \\
  0&1&0
  \end{pmatrix}
  \begin{matrix}
  \bra{-} \\
  \bra{0} \\
  \bra{+}
  \end{matrix}  
 , \quad 
  S_z = \begin{pmatrix}
  0&0&1 \\
  0&0&0\\
  1&0&0
  \end{pmatrix}
  \begin{matrix}
  \bra{-} \\
  \bra{0} \\
  \bra{+}
  \end{matrix}.
  \end{equation}

The symmetry in the $(xy)$ plane is broken by the presence of the transverse electric or magnetic field. The $x_i$ direction is defined as the direction of the transverse electric or magnetic field which dominates the single spin Hamiltonian of the particle $i$.

We will consider two cases :
\begin{itemize}
\item The $x$ direction is defined by an external field (transverse magnetic field or strong external electric field) projected on the $(xy)$ plane of each classes. In particular this means that two spins from the same class have the same $(\hat x, \hat y, \hat z)$ basis.
\item The $x$ direction is defined by the local electric field generated by the charges in the crystal. We will assume that the NV and fluctuator do not see the same electric field, and will sample a random angle $\psi \in [0,2\pi]$ between the axes $\hat{x}_1$ and $\hat{x}_2$.
\end{itemize}

\begin{table}
\begin{tabular}{cccc}
\hline
$\bar{\eta}$ table & $\widehat{z_1 z_2}=0$ & $\widehat{z_1 z_2}=70.5^\circ$ & $\widehat{z_1 z_2}=109.5^\circ$ \\
\hline
$\ket{\pm 1}$ basis & $\frac{2}{3\sqrt{3}}=$ 0.3849 & 0.6507 & 0.8328 \\
$\ket{+/-}$ basis $\hat{x}_1\neq \hat{x}_2$ & 0.7110  & 0.6828 & 0.6828 \\
$\ket{+/-}$ basis $\hat{x}_1= \hat{x}_2$ & $\frac{4}{3\sqrt{3}}=$0.7698  & 0.6951 & 0.6951 \\
\hline
\end{tabular}
\caption{Computation of $\frac{\bar{\eta}}{ \frac{1}{4} \cdot \sqrt{\frac{1}{3}}}$ for the different eigenbasis of the single spin Hamiltonian, and for the different angles between the $z$ axis of the two spins.}
\label{table eta}
\end{table}

For this two cases, which we will refer to as $\hat{x}_1=\hat{x}_2$ and $\hat{x}_1\neq \hat{x}_2$, we can compute the $\bar \eta$ factor in the three scenarii discussed previously : $\widehat{z_1 z_2}=0$ (same class), $\widehat{z_1 z_2}=70.5^\circ$ and $\widehat{z_1 z_2}=109.5^\circ$. The results, as well as those in the $\ket{\pm 1}$ basis are presented in Table \ref{table eta}.

There are two particular situations where these values can be experimentally tested. 

The first situation is described in Fig. 4 in the main text : in the presence of pure transverse magnetic field (for a single class, non-resonant with the three other classes), the decay rate of the class decreases with the magnetic field amplitude, due to the double-flip processes (described below), and then reaches a plateau with a decay rate value $\sim 2$ times larger than in the longitudinal magnetic field case. This increase is in agreement with the fact that $\bar{\eta}_{\rm same}$ is 2 times bigger in the non-magnetic $\ket{+/-}$ basis than in the magnetic $\ket{\pm 1}$ basis. The measured increase is again smaller than the predicted factor of 4 given by eq.~(\ref{eq 1/T}).

The second one is the difference between the $\bm{B} \parallel \langle 100 \rangle$ case and the $\bm{B}=0$ case. When a magnetic field is applied in the [100] direction, all four classes are resonant and the spin Hamiltonian basis is $\{ \ket{0},\ket{+1},\ket{-1} \} $. The $\bar \eta$ factor in this case has been calculated in the previous section : $\bar \eta^2\approx 42.8\, \bar \eta_0^2$. When no external magnetic field is applied, all four classes are also resonant but the spin Hamiltonian basis is  $\{ \ket{0},\ket{+},\ket{-} \} $ (see Sec. \ref{NV Hamiltonian}) and double-flip processes are near-resonant. The decay rate due purely to flip-flop in the $\ket{+/-}$ basis is proportional to $\bar \eta^2 = (\bar{\eta}_{\rm same}^{+/-} + 3\bar{\eta}_{\rm diff}^{+/-})^2 \approx 51.4\, \bar \eta_0^2$. Fig. 3 in the main text shows that we indeed observe a shorter lifetime when $\bm B=0$, in agreement with the higher value of $\bar \eta^2$, but this decrease could also come from the double-flip processes which are absent when $\bm B \neq 0$.

\subsection{Double-flip processes}

Double-flip processes take place where both spins lose or gain one quantum of spin angular momentum, as opposed to flip-flop processes where one spin loses one quantum and the other gains one. These are related to matrix elements  such as $\mel{+1,0}{\frac{\mathcal{H}_{\rm dd}}{J_0/r^3}}{0,-1}$ in the $\ket{\pm 1}$ basis or $\mel{+,0}{\frac{\mathcal{H}_{\rm dd}}{J_0/r^3}}{0,-}$ in the $\ket{\pm}$ basis. These matrix elements couple two-spin states that are never fully resonant, however in small magnetic fields, the residual splitting due to local electric and magnetic field ($\sim 6\ \rm MHz$ with our samples) is small enough compared to the fluctuators line-width  measured in sec. \ref{fluctuator width} to be $\sim 8\ \rm MHz$, so that double-flip processes may still occur.

Fig. 4 in the main text shows that for lower transverse field values, the spin decay rate is increased significantly. We attribute this increase to double-flip processes. We can see that the decay rate increase ($\sim$ 5 times the baseline value) is significantly higher than the increase due to the electric field only in the $\ket{+/-}$ basis ($\sim$ 2 times the baseline value). This lead us to believe that the double-flip processes are the main reason behind the decrease in spin lifetime for $\bm B=0$ observed in Fig. 4 in the main text. 
\bibliographystyle{plain}

\bibliography{Low_field_CR_SI}{}